Research article

# On the propensity of Asn-Gly-containing heptapeptides to form *β*-turn structures : comparison between *ab initio* quantum mechanical calculations and Molecular Dynamics simulations.

Dimitrios A. Mitsikas & Nicholas M. Glykos*

*Department of Molecular Biology and Genetics, Democritus University of Thrace, University campus, 68100 Alexandroupolis, Greece, Tel +30-25510-30620, Fax +30-25510-30620, https://utopia.duth.gr/glykos/ , glykos@mbg.duth.gr*

# Abstract

Both molecular mechanical and quantum mechanical calculations play an important role in describing the behavior and structure of molecules. In this work, we compare for the same peptide systems the results obtained from folding molecular dynamics simulations with previously reported results from quantum mechanical calculations. More specifically, three molecular dynamics simulations of 5 μs each in explicit water solvent were carried out for three Asn-Gly-containing heptapeptides, in order to study their folding and dynamics. Previous data, based on quantum mechanical calculations and the DFT methods have shown that these peptides adopt $β$-turn structures in aqueous solution, with type I' $β$-turn being the most preferred motif. The results from our analyses indicate that for the given system the two methods diverge in their predictions. The possibility of a force field-dependent deficiency is examined as a possible source of the observed discrepancy.

# Running title

Molecular Dynamics simulations of Asn-Gly heptapeptides

## 1. Introduction

$β$-Turns are structural motifs defined by four consecutive residues ($i$ to $i+3$) with the distance between the $C^α(i)$ and $C^α(i+3)$ atoms being less than 7 Å and where the central two residues are not helical (although a $β$-turn may overlap the end of an $α$-helix by up to three residues)[1-5]. They are typically classified into distinct categories based on the distribution of $φ$ and $ψ$ torsion angles of residues $i+1$ and $i+2$[1,3-5]. The almost universally accepted classification scheme is the one proposed by Thornton[1] which defines nine distinct types of $β$-turns: I, I', II, II', VIa1, VIa2, V1b, VIII and IV. Type I and II $β$-turns are the most common types of $β$-turns found in proteins, with the ideal torsion angles ($φ_{i+1}$, $ψ_{i+1}$, $φ_{i+2}$, $ψ_{i+2}$) in each of these categories being (-60°, -30°, -90°, 0°) and (-60°, 120°, 80°, 0°), respectively, whereas, the corresponding values for their mirror-image types I' and II' $β$-turns are (60°, 30°, 90°, 0°) and (60°, -120°, -80°, 0°)[1].

Analyses of X-ray protein structures have provided valuable information about the positional preferences of the 20 amino acids in different types of $β$-turns[1-8]. Residues that show a strong preference for $β$-turns have an important role in stabilizing $β$-hairpin conformations. Studies on the positional preferences of amino acids in positions $i+1$ and $i+2$ of $β$-turns indicated that, among others, Asn is generally favoured in position $i+1$ and Gly in position $i+2$ of type I' and II' $β$-turns, while the same positional preferences are conversed in type I and II $β$-turns[1,4,9,10]. Moreover, analyses of short peptides, revealed that the Asn-Gly segment promotes the formation of type I' $β$-turn and $β$-hairpin conformations though not as effectively as the $^D$Pro-Gly segment[11,12]. Interestingly, thermodynamic analysis has suggested that the entropic advantage of the $^D$Pro-Gly sequence is balanced to some extent by an enthalpic advantage of the Asn-Gly sequence, as the rigid nature of the $^D$Pro-Gly segment may prevent energetically favoured contacts between side-chains[3,12]. However, although the Asn-Gly segment is found frequently in type I' $β$-turns, no extensive computational investigation has been made on the folding of short peptides containing this segment. A recent study by Kang and Yoo, who examined the propensities of three Asn-Gly containing heptapeptides in aqueous solution to form $β$-turn structures using quantum mechanical (QM) calculations and the DFT methods, indicated that type I' β-turn was indeed the preferred structural motif for all the three heptapeptides containing this segment[13]. The optimized torsion angles for every *ab initio* turn type in each heptapeptide, as obtained by DFT calculations, can be seen in Table I. The resulting values are very similar to the ideal values proposed by Thornton's work without large discrepancies, apart from the $\psi$Gly value that despite residing within the ±30° deviation limit, has a slightly increased flexibility compared to the rest of the angles.

In this communication, we attempt to examine the accuracy of Molecular Dynamics (MD) simulations and their ability to reproduce the results derived from experimental data (i.e. Thornton's criteria), as well as *ab initio* calculations. So far, data obtained from classical empirical MD force field analyses regarding the Asn-Gly segment are limited. The most notable finding that gave us cause for reflection on the folding of the Asn-Gly sequence was the inability of the AMBER 99SB*ILDN[14,15] force field to fold and stabilize Trpzip2 —which adopts an Asn-Gly β-turn motif— possibly due to problems describing local conformational preferences[16,17]. Therefore, using the same three Asn-Gly heptapeptides as Kang and Yoo, we decided to study their propensities to adopt β-turn conformations in water, and examine to what extent the MD simulations can provide accurate results concerning protein folding and dynamics.

## 2. Methods

### 2.1 System preparation and simulation protocols

In order to study the propensities of the Asn-Gly segment to form *β*-turn structures we conducted three MD simulations of the same three capped heptapeptides that Kang and Yoo used in their work: Ac-Ala-Ala-**Asn-Gly**-Ala-Ala-NHMe (**hp$_{NG}$-1**), Ac-Leu-Val-**Asn-Gly**-Gln-Tyr-NHMe (**hp$_{NG}$-2**, from PDB entry 1EST) and Ac-Phe-Val-**Asn-Gly**-Leu-Phe-NHMe (**hp$_{NG}$-3**, derived from an octapeptide with the similar sequence Boc-Leu-Phe-Val-Aib-$^D$Ala-Leu-Phe-Val-OMe that forms a type I' Aib-$^D$Ala *β*-turn). The system preparation procedure and simulation protocol has been previously described[18-23] and in summary was performed as follows. Addition of missing hydrogen atoms and solvation-ionization were performed with the LEAP program from the AMBER tools distribution. All three simulations were performed using periodic boundary conditions and a cubic unit cell sufficiently large to guarantee a minimum separation between the PBC-related images of the peptides of at least 16 Å. We followed the dynamics of the peptides' folding simulations using the NAMD program[24,25] for a grand total of 15 μs (5 μs for each peptide), using the TIP3P water model[26], the AMBER 99SB*ILDN force field and adaptive tempering[27] with an inclusive temperature range of 280-480 K as implemented in the NAMD program (adaptive tempering can be considered as a single-copy replica exchange method with a continuous temperature range).

The simulation protocol for all the peptides was the following: the system was first energy minimized for 1000 conjugate gradient steps followed by a heating-up phase to a final temperature of 320 K (with a Δ*T* step of 20 K) over a period of 31 ps. Subsequently, the system was equilibrated for 10 ps under *NpT* conditions without any restraints, until the volume equilibrated. This was followed by the production *NpT* runs with the temperature and pressure controlled using the Nose-Hoover Langevin dynamics and Langevin piston barostat control methods as implemented by the NAMD program, with adaptive tempering applied through the Langevin thermostat, while the pressure was maintained at 1 atm. The Langevin damping coefficient was set to 1 ps$^{-1}$ and the piston oscillation period set to 400 fs with a decay time of 200 fs. The production run was performed with the impulse Verlet-I multiple timestep integration algorithm[28] as implemented by NAMD. The inner timestep was 2.5 fs, with short-range nonbonded interactions being calculated every one timestep, and long-range electrostatic interactions every two timesteps using the particle mesh Ewald method[29] with a grid spacing of approximately 1 Å and a tolerance of 10$^{-6}$. A cutoff for the van der Waals interactions was applied through a switching function, acting between 7 and 9 Å. The SHAKE algorithm[30] with a tolerance of 10$^{-8}$ was used to restrain all bonds involving hydrogen atoms. Trajectories were obtained by saving the atomic coordinates of the whole system every 1 ps.

## 2.2 Trajectory analysis

The programs CARMA[31] and GRCARMA[32] together with custom scripts were used for most of the analyses, including removal of overall rotations/translations, calculation of torsion angles, calculation of the RMSDs from a chosen reference structures, production of PDB files from the trajectories, dihedral space principal component analysis[33-35] and corresponding cluster analysis, etc. Structural analysis was performed using the PROMOTIF[36] and PROCHECK[37] programs. All molecular graphics work and figure preparation were performed using PyMOL[38] and CARMA.

## 2.3 Torsion angle comparison analysis

A significant part of this work focused on analysing the extent to which the *ab initio* $\varphi,\psi$ torsion angles (and the ideal ones defined by Thornton's work) agree with the $\varphi,\psi$ torsion angles derived from the MD simulations. Because the angles from molecular dynamics are continuously variable, this analysis was performed as follows. In the first step we calculated the distances (in degrees) between a reference *ab initio* torsion angle and its corresponding torsion angle from each frame of the trajectory (or from each frame of a dPCA-derived cluster). As reference angles we used the *ab initio* $\varphi$ and $\psi$ torsions of the $X_{-1}$-Asn-Gly-$X_{+1}$ set of residues taking into account every *β*-turn type of each heptapeptide. In the second step, the distribution of these distances was calculated and the resulting histograms plotted. In the final step, we treated these distributions as a sum of independent Gaussian distributions, and we obtained numerical values for the means ($\mu$) and standard deviations ($\sigma$) of those peaks residing closest to the *ab initio* reference value (the calculation of the ($\mu,\sigma$) parameters was performed with the *nlsLM* function of the *R* package[39]). These values were then used to convert distances to probabilities, as will be discussed later.

## 3. Results

### 3.1 Extent of sampling

The heptapeptides —as will be discussed in the next section— demonstrate a highly dynamic behaviour throughout the molecular dynamics simulations. Their structural flexibility/disorder leads to a rugged folding landscape and, consequently, to the absence of a well-defined gradient towards a would-be 'native' structure. The implication of this finding is that the whole configurational space available to the peptides is accessible during the simulations. This immediately raises the issue of convergence and statistical significance of our results. To tackle this problem, we apply a recently proposed probabilistic method for estimating the convergence of molecular dynamics trajectories. The method is based on the application of Good-Turing statistics to estimate the amount of structural variability that has *not* been observed in a given trajectory. These probabilistic estimates are presented in the form of graphs that depict the probability of observing a new/different structure, as a function of the minimal RMSD —of this (thus far unobserved) structure— from all those structures already observed[40]. The results from the application of Good-Turing statistics are shown in Figure 1.

To clarify these graphs, the high probability of unobserved species for low RMSD values indicates that if we were to continue the simulation (even for just one additional timestep) it would be highly probable to observe structures, which although similar to some of the already observed ones, they would still be slightly different from them. As the RMSD from the already observed structures increases, the corresponding probability decreases. It is therefore the exact form of this graph and the inclination of each curve towards low probability values that inform us to what extent the trajectories have been sampled or, in other words, how significant is the structural variability that has been missed due to limited sampling. Focusing on the application of Good-Turing statistics in our trajectories, the results have been obtained from multiple calculations and are organised in two groups. Figure 1 illustrates the direct application of the method to the MD trajectories considering only the backbone atoms for either the entire heptapeptides (upper three curves) or their four-residue central part (lower three curves). For the entire peptides, the statistical analysis implies that if we continued the simulations, we would expect that on average 1/10 of the new (previously unobserved) structures would differ by an RMSD of at least ~1.3 Å. Clearly, the effect of limiting the residue selection to the amino acids forming the *β*-turn is rather dramatic: the curves fall quite fast to negligible $P_{unobserved}$ values for RMSDs of the order of 1 Å. The apparent difference between these two cases may be attributed to the fact that the four-residue central parts are structurally more stable, possibly promoting the formation of secondary-structure-like patterns, whereas the peptides' termini are highly mobile and disordered.

To conclude, the results drawn from the analysis of our 5 μs trajectories indicate that the length of the simulations is probably sufficient and guarantees a reasonable sampling of the peptides' configurational space for the given force field. A similar type of analysis has also been made but using only structures whose corresponding adaptive tempering temperature was ≤ 360 K, thus corresponding to more stable (from the simulation's point of view) conformers. As shown in Figure S1 of the Supplementary Information file, the differences are minor compared to the first analysis and setting a temperature cutoff for examining only more stable peptide conformers does not result in our case to a better sampling of the configurational space.

**3.2 The peptides are very flexible, but with a tendency to form *β*-turns**

To place our observations on a structurally firm ground, we calculated the free energy landscapes of the heptapeptides using the dihedral angle Principal Component Analysis (dPCA) method to identify the prominent conformations. Various studies have already indicated that dPCA is a high resolution, powerful and more appropriate method for studying such highly flexible systems[27-29]. We restricted our analysis to the backbone dihedral angles of the four central residues only, since the rest of the peptides were found to be kinetically frustrated and uncorrelated to the rest of the system. The results from the dPCA analysis are shown in Figure 2 in the form of a set of diagrams illustrating the log density projections of the corresponding trajectories along the top two principal components (i.e. the free energy landscapes). High density peaks (dark blue regions) can be associated with distinct peptide conformers. Representative structures of these clusters were extracted after calculating an average structure of each cluster and then selecting the frame of the trajectory with the lowest RMS deviation from the corresponding average structure. Representative structures corresponding to *β*-turn motifs for each heptapeptide can be seen in Figure 3. Note that the actual cluster analysis was performed in the three-dimensional principal component space, and not in two dimensions as shown here for clarity.

As it can be seen in Figure 2, the free energy landscapes are quite rugged with several free energy minima that correspond to numerous distinct conformations, rather than a few stable and persistent structures. For **hp$_{NG}$-1**, 34.7% of the total frames were assigned by the program CARMA to clusters, while for **hp$_{NG}$-2** and **hp$_{NG}$-3** the corresponding percentages were 36.7% and 31.8%, respectively. The most highly populated cluster recorded for each heptapeptide occupied only 11.9%, 20.1% and 20.7% of the clustered frames (for the **hp$_{NG}$-1**, **hp$_{NG}$-2** and **hp$_{NG}$-3** respectively), clearly indicating that the peptides demonstrate a very dynamic behaviour. Table SI in Supplementary Information file offers a detailed view regarding the population (among clustered frames) of each cluster and the occupancy of *β*-turn motifs in each cluster. There is a structural variability concerning the preference among turn types and the positional preference of each amino acid in positions *i*+1 and *i*+2. Asn-Gly turns, not only failed to be the dominant structural motif, but the analysis revealed even the presence of a plethora of *β*-turn motifs as well. Examination of clusters' representative structures reveals that structures can vary among different *β*-turn conformations. Figure 3 shows a collection of those clusters' representative structures that form a *β*-turn, clearly demonstrating the large diversity of the turn types that have been sampled. Many representative structures corresponding to significantly populated clusters adopt a *β*-turn conformation, but the presence of unfolded random coil representatives is nonetheless considerable. Asn-Gly is not favoured to be the turn segment among the representatives, as there is a variety in the positional preference of each residue for this site.

To have a more comprehensive understanding of the preferences among the different $\beta$-turn motifs, Table II shows the populations of $\beta$-turns sampled in each trajectory taking into account both the different combinations of $i+1$ and $i+2$ positional preferences and the different types of $\beta$-turns. Of the turns located, $\beta$IV turns seem to be the most prominent turn motif in each trajectory, but this can be attributed to the fact that this type does not follow any of the stringent criteria being used for turns' classification scheme. $\beta$VIII turns on the other hand, were poorly sampled throughout the simulations resulting only in a negligible amount of observations. As for the most common turn motifs $\beta$I, $\beta$I', $\beta$II, $\beta$II', $X_{-1}$-Asn and Gly-$X_{+1}$ $\beta$I turns appears to be the most frequently sampled turn motifs in each trajectory followed by Gly-Ala $\beta$II' turns in **hp$_{NG}$-1**, Gly-Gln $\beta$II' turns in **hp$_{NG}$-2** and Asn-Gly $\beta$II turns in **hp$_{NG}$-3**. The Asn-Gly segment is slightly to moderately favoured among these common turn types depending on the heptapeptide, showing a rather increased preference towards $\beta$I turns in **hp$_{NG}$-1** and $\beta$I', $\beta$II turns in **hp$_{NG}$-2** and **hp$_{NG}$-3**. Regarding type II $\beta$-turns, although their sampling is limited, Gly-$X_{+1}$ shows a relatively increased preference as a turn segment for this turn type. These results are in good agreement with previous studies[1,5] on the positional preferences of amino-acids in turns indicating the strong preference of Gly in positions $i+3$ in $\beta$I turns, $i+2$ in $\beta$I' and $\beta$II turns and $i+1$ in $\beta$II' turns, as well as of Asn in position $i+1$ in $\beta$I' turns. Our findings, however, do not agree with the QM-DFT calculations of Kang and Yoo. As stated in their work, the most favoured conformation for all three heptapeptides according to their calculations is the Asn-Gly $\beta$I' turn motif (for **hp$_{NG}$-1** $\beta$II' and $\beta$I' turn motifs were almost equally preferred). In contrast, the present analysis shows that Asn-Gly turns are overall (Table II) the least sampled turn motif for the MD simulations with their populations reaching ~20-25% of the total frames in each trajectory, contrary to those of $X_{-1}$-Asn and Gly-$X_{+1}$ turns that correspond to ~30% of total frames. Overall, the analysis of the molecular dynamics simulations indicates that although the presence of $\beta$-turn motifs is a very significant feature of the peptides' structural behaviour, the trajectories are also characterized by a significant structural diversity, rather than a dominant presence of only type I' $\beta$-turns or an exclusive preference for the Asn-Gly sequence as the turn segment.

**3.3 Torsion angles derived from MD generally agree with the *ab initio* and the ideal *β*-turn *φ,ψ* values, but significant discrepancies are present**

In the previous section we examined the general structural characteristics of the simulations in terms of preferred peptide conformers and turn types. In this section we make a more detailed comparison with the *ab initio* results through a direct numerical evaluation of the differences observed for individual torsion angles. The procedure is based on calculating for each frame (and for each of the trajectories) the torsion angles of the central four-residue part of the heptapeptides, followed by the calculation of the deviation (in degrees) between the MD-derived angles with the *ab initio* ones. Figure 4 shows the results in the form of histograms illustrating the deviation of each torsion angle from the corresponding *ab initio* derived one. Values near $d=0°$ indicate similarity with the *ab initio* model (as the differences between the values of the MD torsion angle and the corresponding *ab initio* are small), while values far from $d=0°$ indicate large differences. To further quantify these observations, we located the Gaussian distributions closest to the *ab initio* values (i.e. the $d=0°$) and we obtained numerical values for their means ($\mu$) and standard deviations ($\sigma$) through a nonlinear regression fitting performed with the *nlsLM* function of the *R* package (red curves in Figure 4). Using these values we then calculated Z-scores and the corresponding probabilities for the observed deviations between the *ab initio* peptide structures and the molecular dynamics-derived ones. The numerical results are shown in Table SII of the Supplementary Information file. In this table, large Z-scores (and their associated low probabilities) signify a statistically important deviation between *ab initio* and MD. Overall the agreement between the two methods appears to be reasonable but with some notable exceptions. On one hand, almost half of the MD-derived values lie within one standard deviation from the *ab initio* reference angle values, while a significant number of them lie within two standard deviations. This implies that the majority of the *ab initio* angle values have also been sampled by the molecular dynamics simulations giving rise to local conformations that are similar to the QM-derived *β*-turn structures. On the other hand, there are angles exhibiting large deviations from the *ab initio* values. As we can see in Figure 4, while there are torsions sampled in the trajectories similar to the *ab initio* angles (Gaussian distributions in the middle of panels

A-B of Figure 4), in some cases examined, the 99SB*ILDN force field failed not only to reproduce, but even to sample, otherwise allowed values (as denoted by the structural evaluation using the PROCHECK program in Figures S2-S4 of Supplementary Information file). Such examples are shown in panels C-H of Figure 4. The most obvious and persistent deficiency is the inability of the force field to successfully fold *β*II' Asn-Gly structures (Figure 4 C-E). Previous data[17] however have already shown that the 99SB*-ILDN force field fails to adequately fold and stabilize the Trpzip-2 peptide which also adopts an Asn-Gly *β*-hairpin conformation[16,17], giving only a small fraction of folded structures under equilibrium conditions, possibly due to problems describing local conformational preferences. In particular, folding simulations of this peptide using the 99SB*-ILDN force field, showed that the *β*-hairpin of Trpzip-2 formation could not be stabilized, as well as that there is an increase in *a*-helix propensity due to an imbalance of the local conformational preferences. This suggests the probability of a force field bias towards *α*-helical conformations, and therefore of an imbalance on the intrinsic conformational preferences, thus making the stabilization of the *β*-hairpin unachievable.[17]

Table III shows an overview of all torsion angles for which their mean values as derived from molecular dynamics deviate by more than 30° from their corresponding *ab initio* values. Despite the fact that there are obvious, but nonrecurring, discrepancies in some angle values between the MD and the *ab initio* conformations (except those of *ψ*Asn *β*II' values, which are persistent in each peptide system), the angle values converge to some extent with the QM-DFT structures. To aid visualization of these findings, we show in Figure 5 the *ab initio* structures but colored according to the magnitude of deviation between the *ab initio* angles and the simulation derived ones. The colour coding in Figure 5 is indicative of the MD-*ab initio* deviations and ranges from blue (small deviations) via green (moderate deviations) to red (large deviations). Torsion angles that demonstrate the largest deviations throughout the simulations are shown labelled. It should be noted here that over all peptides and turn types, there is only one angle that demonstrates a persistent and consistent deviation, and this is the *ψ*Asn angle for the *β*II' conformation. Closer examination of Figure 4 and Table III shows that the distributions of *ψ*Asn values from the simulations cluster around three distinct regions. The large peak [right-hand side peak in panels

C-E of Figure 4] corresponds to an area (in the Ramachandran plot) centered around $\psi \sim +145°$ while the two shorter peaks [left-hand side of panels C-E of Figure 4] correspond to areas centered on $\psi \sim +15°$ and $\psi \sim -25°$. These areas are close to the proposed values of $\psi = +120°$ for the position $i+1$ in $\beta$II turns, $\psi = \pm 30°$ for the position $i+1$ in $\beta$I and $\beta$I' turns and $\psi = 0°$ for the $i+2$ positions of all $\beta$-turn types. These results not only agree, but also explain both the presence of significant populations of $X_{-1}$-Asn turns and Asn-Gly $\beta$I, $\beta$I' and $\beta$II turns, as well as the almost complete absence of Asn-Gly $\beta$II' turns from the trajectories.

To complete this set of comparisons, we also calculated the distances (in degrees) between the Asn-Gly torsion angles from molecular dynamics with the corresponding $\varphi,\psi$ $\beta$-turn values from Thornton's definitions. The results are shown in Table IV and agree and reinforce the conclusions drawn from the comparison with the *ab initio* values : most of the torsion angles from the trajectories fluctuate around the ideal $\varphi,\psi$ values (with their mean values residing within the deviation limit of $\pm 30°$). The outliers are again $\psi$Asn in $\beta$II' **hp$_{NG}$-1**, $\psi$Asn in $\beta$II' **hp$_{NG}$-2** and $\psi$Asn in $\beta$II' **hp$_{NG}$-3** which show a very significant deviation from the ideal $\beta$-turn values, far exceeding the $\pm 30°$ limits set by Thornton and coworkers.

### 3.4 Torsion angle analysis of the dPCA-derived clusters confirms and reinforces the results from the whole trajectories

To compare our previous observations with the results obtained from the dPCA, we calculated as described above the distance of the $\varphi, \psi$ angles in each cluster's frame from the corresponding *ab initio* value. Figures S5-S7 of the Supporting Information file show —for the five most populated clusters of each heptapeptide— the distributions of these distances for the torsion angles of the four-residue central part together with the occupancy of the corresponding $\beta$-turns and a representative structure for each cluster. A closer look at the histograms reveals a significant divergence of the Asn-Gly $\varphi,\psi$ peaks from the $d=0°$ value for several clusters. The cases where this torsion angle analysis indicates convergence with the *ab initio* data can also be verified through a comparison with the results shown in Table SI. Clusters 09 and 03 of **hp$_{NG}$-1**, for example, have significant populations of Asn-Gly type II and I $\beta$-turns respectively as the sharp Asn-Gly $\varphi$, $\psi$ peaks lying around $d=0°$ indicate (Figure S5). Clusters 01 and 02 of **hp$_{NG}$-2** are occupied mainly by Asn-Gly $\beta$I' and $\beta$I turns respectively (Figure S6), whereas for **hp$_{NG}$-3**, clusters 01, 03 and 07 are heavily populated by Asn-Gly $\beta$I, $\beta$I' and $\beta$II turns respectively (Figure S7). Figure S8 of the Supporting Information file shows in the form of barcharts the occupancy of Asn-Gly turns for each trajectory and cluster based on a PROMOTIF analysis of a sample of 500 randomly selected structures derived from each dPCA cluster. Bars are coloured according to the type of different $\beta$-turn motifs, and indicate the percentage (out of these 500 structures) of Asn-Gly turns sampled in the clusters. The results are in good agreement with the previously observed populations of Table SI and Figures S5-S7. Asn-Gly turns are indeed prefered to some extent by our peptide systems, but they are not the major structural motifs observed. Some of the major clusters are not even occupied by Asn-Gly turns indicating that many Asn-Gly turn conformers may have been sampled throughout the trajectory and do not belong in any of the dPCA-derived clusters. Similarly, the Asn-Gly $\beta$I' turn although it is present, it is definitely not a predominant structural motif of the dPCA clusters. A question that arises at this point is this : Can the clusters be characterized by a single structural motif ? In other words, can a highly populated motif be representative of a cluster's structural ensemble ? If we look in Figures S5-S7, even though there are clusters with significant populations of a certain $\beta$-turn type, they cannot be unequivocally identified by a single motif. As can be seen in Figures S5-S7, representative structures differ from the general structural propensities that are apparent in each cluster. Structural analysis of the representative conformations shown in Figure 3 also indicated that representative structures may adopt conformations that differ significantly from the heavily populated structures identified in each cluster. In a sense, this variability is not unexpected : the analysis in Figures S5-S7 attempts to characterize the structural motifs observed in the whole of the heptapeptides but based on the clustering obtained from a dPCA analysis of only the four central residues. The reason for our choice to focus the analysis on the four central residues is, of course, that is simplifies the comparison with the *ab initio* data which is the principal target of this investigation.

## 4. Summary and Conclusions

The primary aim of this communication was to evaluate the ability of MD simulations to reproduce the structure of turn-forming peptides whose structures were previously characterized from QM calculations. The *ab initio* calculations indicated that the type I' *β*-turn is the most preferred motif in aqueous solution for the three heptapeptides containing the Asn-Gly segment (with some variations depending on the sequence of the peptides and the solvent polarity). The picture painted by molecular dynamics, however, is fundamentally different : according to the simulations, the peptides are highly flexible with multiple relatively shallow free energy minima that allow the peptides to quickly interconvert between structurally diverse conformations. The dPCA-based free energy landscapes in Figure 2 and the corresponding cluster analysis show that there is no funnel-like gradient leading to a native state and, thus, there are no prominent and persistent highly populated structures. The apparent higher stability of the peptides' central four-residue part does not invalidate the general conclusion about the highly dynamic behaviour of these systems. Having noted those differences between QM and MD, we must not fail to also stress their fundamental similarity : these peptides do show a very strong preference for *β*-turn formation. As Table II shows, the various types of *β*-turns containing Asn/Gly as one of their central residues occupy the largest part of the respective MD trajectories with cumulative frequencies reaching ~85% for two of the peptides (becoming ~45% if the type IV turns are excluded). The residue-specific preferences for positions $i+1$ and $i+2$ on the *β*-turns varies significantly between the four central residues, with the Asn-Gly segment preferring the aforementioned positions almost equally with segments formed by neighboring residues ($X_{-1}$-Asn and Gly-$X_{+1}$). Type I *β*-turns appear to be the most prominent MD-derived motif in each trajectory, while Asn-Gly *β*I' turns show significant populations only for the **hp$_{NG}$-2** and **hp$_{NG}$-3** peptides. In agreement with the dPCA results, the torsion angle analysis also demonstrated the structural malleability of the peptides and their pronounced tendency to form various types of *β*-turns. But it was also this more detailed analysis that allowed us to convincingly demonstrate the inability of the the ff99SB*-ILDN force field to sample the Asn-Gly *β*II' motif. This apparent force field bias was also confirmed through a comparison between MD and the ideal *β*-turn values obtained by the Thornton group. This comparison (shown in Table IV) showed a very reasonable agreement between the two sets with the pronounced exception of the $\psi$Asn angles for the $\beta$II' turns. Given the long and successful history of applications of the AMBER 99SB family of force fields to peptide folding, we decided to perform one additional simulation of only the **hp$_{NG}$-1** peptide but this time using the 99SB-ILDN force field instead of the ff99SB*-ILDN variant. The aim of this additional simulation was to establish whether the apparent force field deficiency concerning mirror-image $\beta$-turns was introduced in the force field during its optimization to better sample $\alpha$-helical structures. Analysis of this 99SB-ILDN trajectory with the PROMOTIF program and comparison with the results shown in Table II clearly indicated that this is not the case : the cumulative percentages for the I' and II' turns remain virtually unchanged between the two force fields. To put this in numbers, the observed changes for the various I' and II' turns (compared with the **hp$_{NG}$-1** entries in Table II), are : **I'** : 0.41→0.47%, 2.34→2.54%, 0.38→0.31%, **II'** : 0.17→0.08%, 0.04→0.03%, 7.35→6.75%.

Other than the structural and computational interest on the peptides' structural preferences and the corresponding comparison between the MD and QM-DFT methods, the work reported here has clear implications concerning the validation of the force fields, and especially the number of case studies and computational effort required to discover any yet unidentified problems with new force fields. An inescapable conclusion from this work is that just because a force field has successfully been used to study the folding of a large number of peptides, this alone neither implies nor guarantees the absence of systematic errors that would require the application of very specific structural motifs and peptides for these errors to become obvious. Having noted this conclusion, we feel that we should also argue against it : our analyses and comparisons were not performed against solid experimental data (obtained, for example, from an NMR study of the peptides), and can thus be argued that we could be completely wrong in assuming that these specific peptides should actually sample, for example, the $\beta$II' turn conformation. Having said that, the amount of publications on the Asn-Gly containing peptides, and especially previous NMR and X-ray studies[41-43] that have shown the preference of the Asn-Gly sequence for "mirror-image" $\beta$-turns (type I' or type II'), leaves little doubt that what we have here is a case where a minor force field bias precludes the formation of the experimentally expected structural motifs. Seeing our results in a negative light, it could be argued that this is a clear reminder that as the pace of producing and publishing new molecular dynamics force fields is increasing, the probability that any of these force fields can be reasonably validated with a multitude of structurally independent case studies is correspondingly decreasing, to the point of making the choice of a suitable force field an exercise in futility.

# Figure Captions

**Figure 1.** Extent of sampling and statistical significance. Results from the application of Good-Turing statistics to the three trajectories for both the full-length peptides and their four-residue central part. See text for details.

**Figure 2.** Diagrams illustrating the free energy landscape of each heptapeptide along the top two principal components, as obtained by the dihedral angle Principal Component Analysis (blue peaks correspond to high density regions). Shown with numbers are the different clusters derived by the dPCA analysis.

**Figure 3.** Cluster's backbone plus $C^\beta$ representative structures that correspond to *β*-turn motifs. The structural analysis was performed using the PROMOTIF program and structures were created using the PyMOL program. The colour coding used here is representative of the atoms' type (blue colour corresponds to N atoms, red colour corresponds to O atoms and light grey colour corresponds to C atoms). Shown are only representative *β*-turn structures that correspond to clusters of population > 1%.

**Figure 4.** Histograms showing the distribution of distances (in °) between the *ab initio* $\varphi,\psi$ values and the $\varphi,\psi$ values obtained from the MD simulations for selected residues. While there are torsion angles of specific residues that fluctuate around the *ab initio* values ($d=0°$) during the simulation (A-B), some torsions present a significant divergence from the proposed values (C-H).

**Figure 5.** Shown are the *ab initio* structures illustrating the torsion angle discrepancies in relation with the MD derived angles. The colour coding is indicative of the mean distance of the MD $\varphi,\psi$ values from the corresponding *ab initio* ones, varying from blue (small deviations) to green (moderate deviations) and red (large deviations). Shown with a label, are the torsion angles presenting the greatest deviation through the simulations.

**Table I.** Torsion angles (°) of β-turn residues as obtained from the *ab initio* structures

| Peptide | Turn type | $\varphi_i$ | $\psi_i$ | $\varphi_{i+1}$ | $\psi_{i+1}$ | $\varphi_{i+2}$ | $\psi_{i+2}$ | $\varphi_{i+3}$ | $\psi_{i+3}$ |
|---|---|---|---|---|---|---|---|---|---|
| hp$_{NG}$-1 | I | -64.5 | 165.4 | -54.5 | -37.0 | -86.6 | -12.3 | -170.0 | 148.9 |
|  | I' | -86.0 | 92.2 | 53.3 | 41.5 | 98.7 | -16.5 | -77.8 | 92.5 |
|  | II | -62.1 | 163.9 | -50.5 | 134.9 | 77.3 | -6.9 | -164.7 | 129.5 |
|  | II' | -84.3 | 96.0 | 51.9 | -133.5 | -71.3 | -9.8 | -80.5 | 90.2 |
| hp$_{NG}$-2 | I | -106.0 | 153.7 | -54.5 | -35.1 | -74.9 | -26.2 | -175.5 | 150.4 |
|  | I' | -132.4 | 119.6 | 57.8 | 36.1 | 91.2 | -23.5 | -80.0 | 109.3 |
|  | II | -108.6 | 151.0 | -47.8 | 130.4 | 89.9 | -18.1 | -166.6 | 128.0 |
|  | II' | -130.1 | 132.5 | 57.1 | -134.9 | -87.8 | -7.4 | -84.1 | 108.9 |
| hp$_{NG}$-3 | I | -130.0 | 168.9 | -55.1 | -28.3 | -112.2 | 25.8 | -158.4 | 145.3 |
|  | I' | -145.3 | 121.2 | 58.3 | 38.7 | 78.1 | -4.8 | -106.4 | -169.9 |
|  | II | -129.7 | 167.6 | -47.3 | 136.6 | 63.9 | 23.0 | -160.0 | 130.6 |
|  | II' | -78.9 | 85.1 | 54.2 | -146.9 | -65.9 | -24.8 | -56.1 | 150.2 |

**Table II.** Populations (%) of the various β-turn types sampled in the MD trajectories[a]

| β-Turn type[b] | hp$_{NG}$-1 | | | hp$_{NG}$-2 | | | hp$_{NG}$-3 | | |
|---|---|---|---|---|---|---|---|---|---|
| | Ala-Asn | Asn-Gly | Gly-Ala | Val-Asn | Asn-Gly | Gly-Gln | Val-Asn | Asn-Gly | Gly-Leu |
| I | 9.71 | 3.20 | 8.16 | 10.73 | 3.88 | 9.04 | 12.46 | 4.60 | 8.86 |
| I' | 0.41 | 2.34 | 0.38 | 0.28 | 6.22 | 0.19 | 0.24 | 5.64 | 0.17 |
| II | 0.62 | 5.36 | 0.81 | 0.75 | 5.55 | 0.31 | 0.89 | 6.19 | 0.28 |
| II' | 0.17 | 0.04 | 7.35 | 0.06 | 0.03 | 6.35 | 0.06 | 0.03 | 3.70 |
| VIII | 1.51 | 0.13 | 0.40 | 1.07 | 0.10 | 0.43 | 0.96 | 0.12 | 0.36 |
| IV | 12.47 | 8.12 | 13.12 | 16.24 | 9.19 | 17.18 | 15.17 | 8.95 | 17.22 |
| Total | 24.89 | 19.19 | 30.22 | 29.12 | 24.97 | 33.49 | 29.78 | 25.54 | 30.58 |

[a]Shown are the % trajectory populations of the different β-turns containing Asn/Gly as one of their central residues. The assignment of β-turns was performed using the PROMOTIF program.

[b]Types VIa1, VIa2 and VIb are excluded from our analysis as they require a Pro residue in position $i+2$.

Table III

**Table III.** Torsion angles exceeding the ±30° deviation limit from the *ab initio* angles, along with their corresponding distances in terms of absolute value

| hp$_{NG}$-1 | hp$_{NG}$-2 | hp$_{NG}$-3 |
|---|---|---|
| $\varphi$Ala$_2$ (\|39.6°\|) $\beta$**I** | $\varphi$Gln (\|45.9°\|) $\beta$**I** | $\varphi$Gly (\|41.2°\|) $\beta$**I** |
| $\psi$Ala$_1$ (\|62.3°\|) $\beta$**I'** | $\psi$Gln (\|38.4°\|) $\beta$**I'** | $\varphi$Leu (\|38.6°\|) $\beta$**I** |
| $\psi$Asn (\|33.0°\|) $\beta$**I'** | $\varphi$Gln (\|36.8°\|) $\beta$**II** | $\psi$Leu (\|42.2°\|) $\beta$**I'** |
| $\psi$Ala$_2$ (\|61.6°\|) $\beta$**I'** | $\psi$Asn (\|77.9°\|) $\beta$**II'** | $\varphi$Leu (\|40.2°\|) $\beta$**II** |
| $\varphi$Ala$_2$ (\|34.3°\|) $\beta$**II** | $\psi$Gln (\|38.8°\|) $\beta$**II'** | $\psi$Val (\|58.4°\|) $\beta$**II'** |
| $\psi$Ala$_1$ (\|58.5°\|) $\beta$**II'** | | $\psi$Asn (\|67.6°\|) $\beta$**II'** |
| $\psi$Asn (\|80.3°\|) $\beta$**II'** | | |
| $\psi$Ala$_2$ (\|81.9°\|) $\beta$**II'** | | |

**Table IV**

**Table IV.** Absolute differences between the calculated Gaussian mean values and the ideal $\varphi$, $\psi$ torsion angles, for Asn and Gly residues in each $\beta$-turn type.

| Turn type | hp$_{NG}$-1 | hp$_{NG}$-2 | hp$_{NG}$-3 |
|---|---|---|---|
| $\beta$I | $\varphi$Asn (\|14.9°\|) | $\varphi$Asn (\|13.1°\|) | $\varphi$Asn (\|12.1°\|) |
| | $\psi$Asn (\|2.1°\|) | $\psi$Asn (\|3.3°\|) | $\psi$Asn (\|0.4°\|) |
| | $\varphi$Gly (\|18.2°\|) | $\varphi$Gly (\|19.5°\|) | $\varphi$Gly (\|19.0°\|) |
| | $\psi$Gly (\|5.4°\|) | $\psi$Gly (\|1.9°\|) | $\psi$Gly (\|3.0°\|) |
| $\beta$I' | $\varphi$Asn (\|28.3°\|) | $\varphi$Asn (\|7.8°\|) | $\varphi$Asn (\|7.2°\|) |
| | $\psi$Asn (\|21.5°\|) | $\psi$Asn (\|9.3°\|) | $\psi$Asn (\|19.9°\|) |
| | $\varphi$Gly (\|17.7°\|) | $\varphi$Gly (\|18.1°\|) | $\varphi$Gly (\|15.8°\|) |
| | $\psi$Gly (\|5.4°\|) | $\psi$Gly (\|1.9°\|) | $\psi$Gly (\|3.0°\|) |
| $\beta$II | $\varphi$Asn (\|14.9°\|) | $\varphi$Asn (\|13.1°\|) | $\varphi$Asn (\|12.1°\|) |
| | $\psi$Asn (\|26.0°\|) | $\psi$Asn (\|26.9°\|) | $\psi$Asn (\|25.6°\|) |
| | $\varphi$Gly (\|8.1°\|) | $\varphi$Gly (\|8.1°\|) | $\varphi$Gly (\|5.8°\|) |
| | $\psi$Gly (\|5.4°\|) | $\psi$Gly (\|1.9°\|) | $\psi$Gly (\|3.0°\|) |
| $\beta$II' | $\varphi$Asn (\|31.1°\|) | $\varphi$Asn (\|7.8°\|) | $\varphi$Asn (\|7.2°\|) |
| | <span style="color:red">$\psi$Asn (\|93.1°\|)</span> | <span style="color:red">$\psi$Asn (\|92.8°\|)</span> | <span style="color:red">$\psi$Asn (\|94.51°\|)</span> |
| | $\varphi$Gly (\|8.2°\|) | $\varphi$Gly (\|9.5°\|) | $\varphi$Gly (\|8.9°\|) |
| | $\psi$Gly (\|5.4°\|) | $\psi$Gly (\|1.9°\|) | $\psi$Gly (\|3.0°\|) |

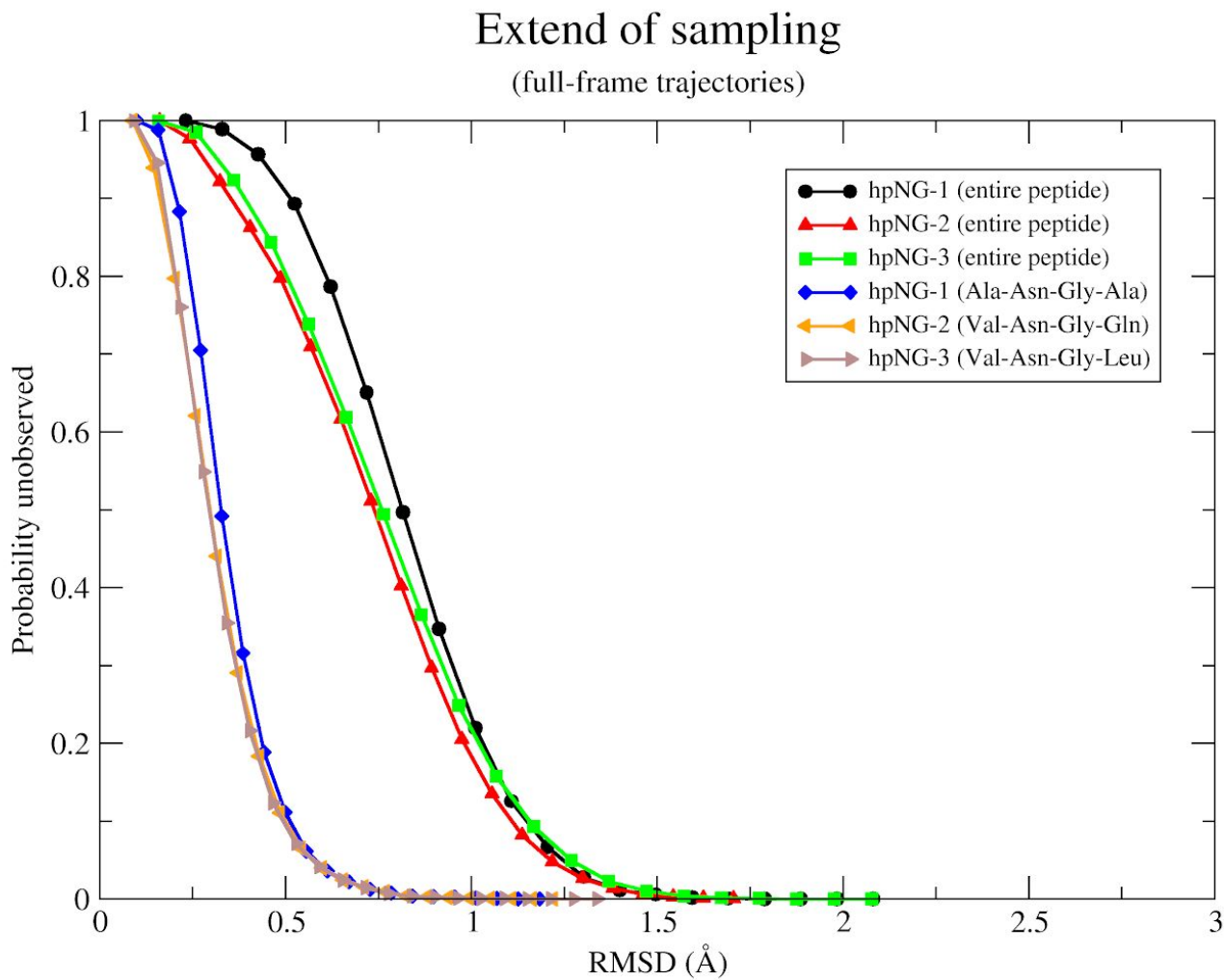

**Figure 1.** Extent of sampling and statistical significance. Results from the application of Good-Turing statistics to the three trajectories for both the full-length peptides and their four-residue central part. See text for details.

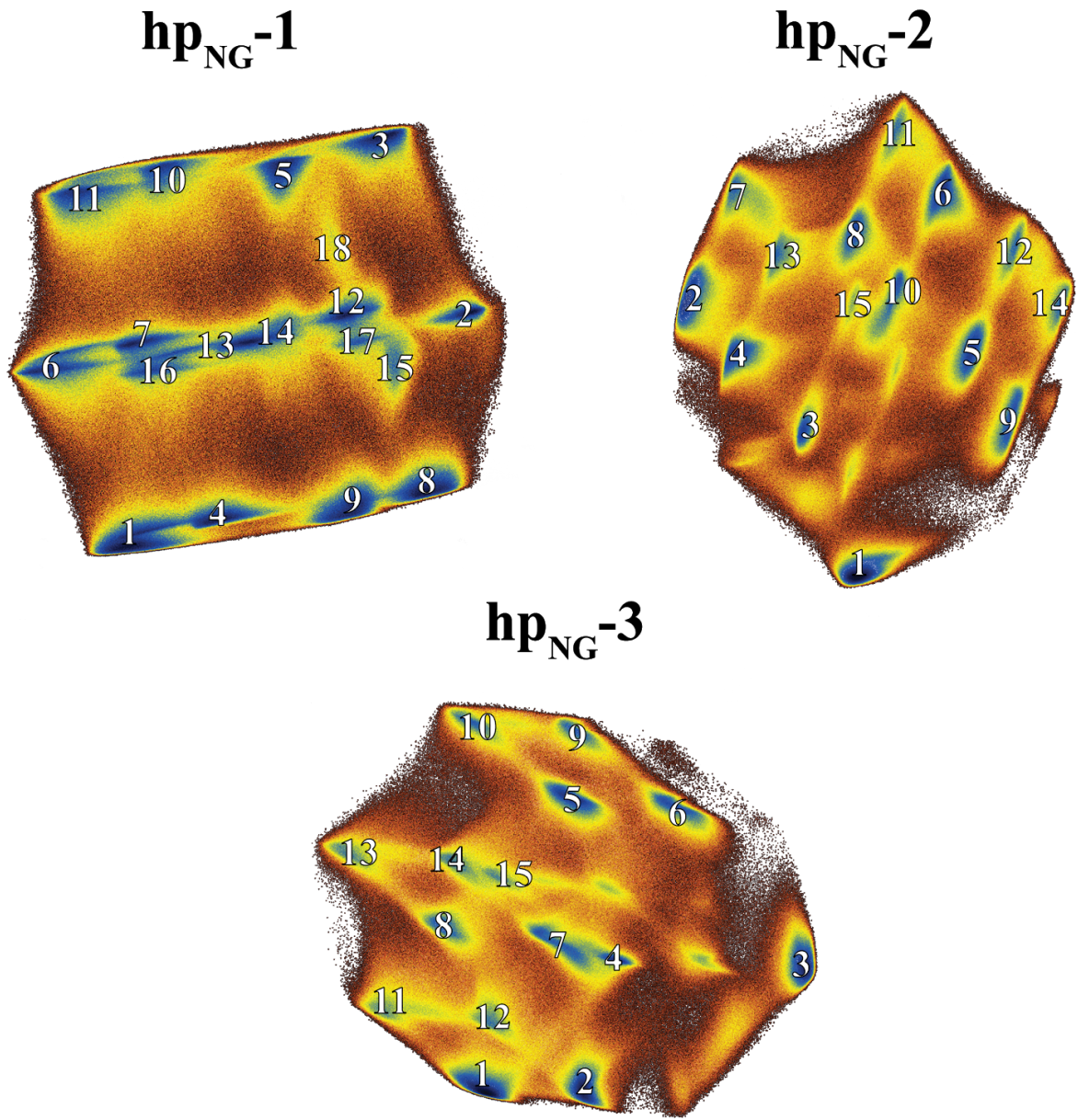

**Figure 2.** Diagrams illustrating the free energy landscape of each heptapeptide along the top two principal components, as obtained by the dihedral angle Principal Component Analysis (blue peaks correspond to high density regions). Shown with numbers are the different clusters derived by the dPCA analysis.

**Figure 3**

## hp$_{NG}$-1

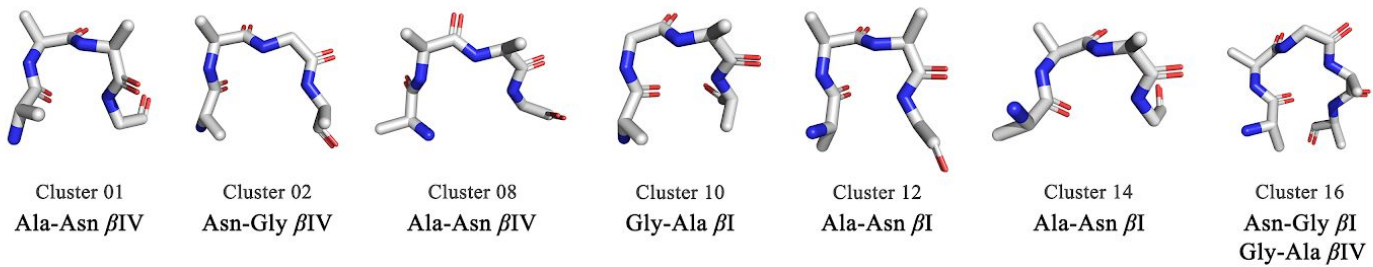

## hp$_{NG}$-2

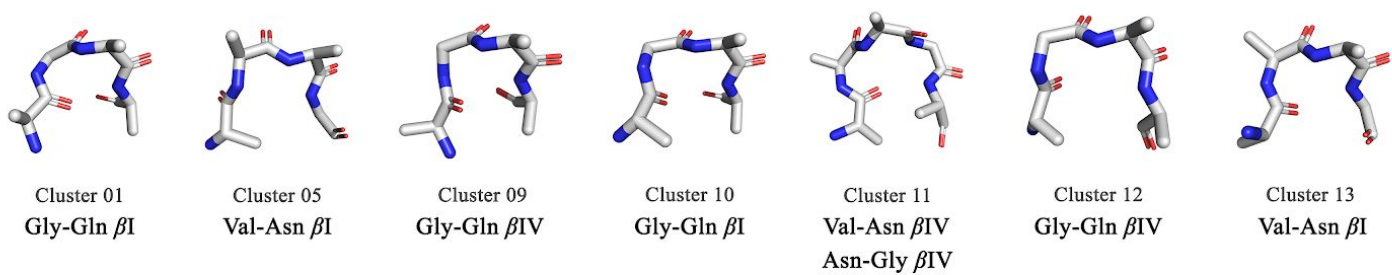

## hp$_{NG}$-3

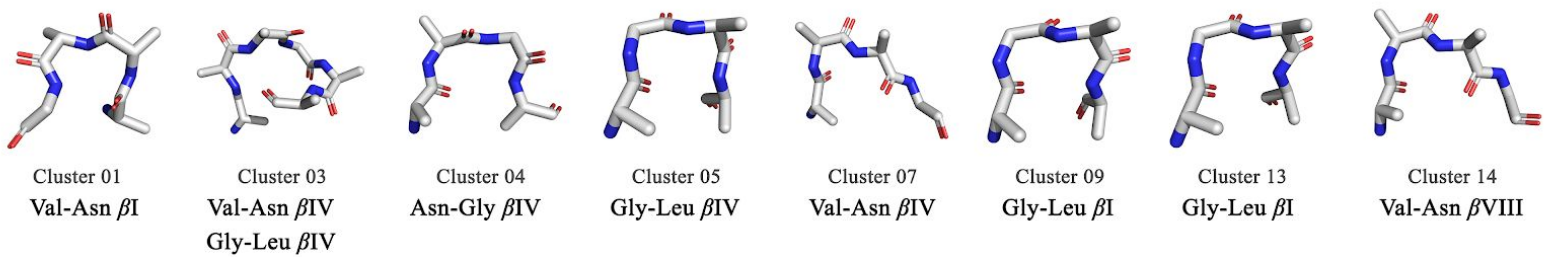

**Figure 3.** Cluster's backbone plus C$^{\beta}$ representative structures that correspond to *β*-turn motifs. The structural analysis was performed using the PROMOTIF program and structures were created using the PyMOL program. The colour coding used here is representative of the atoms' type (blue colour corresponds to N atoms, red colour corresponds to O atoms and light grey colour corresponds to C atoms). Shown are only representative *β*-turn structures that correspond to clusters of population > 1%.

**Figure 4**

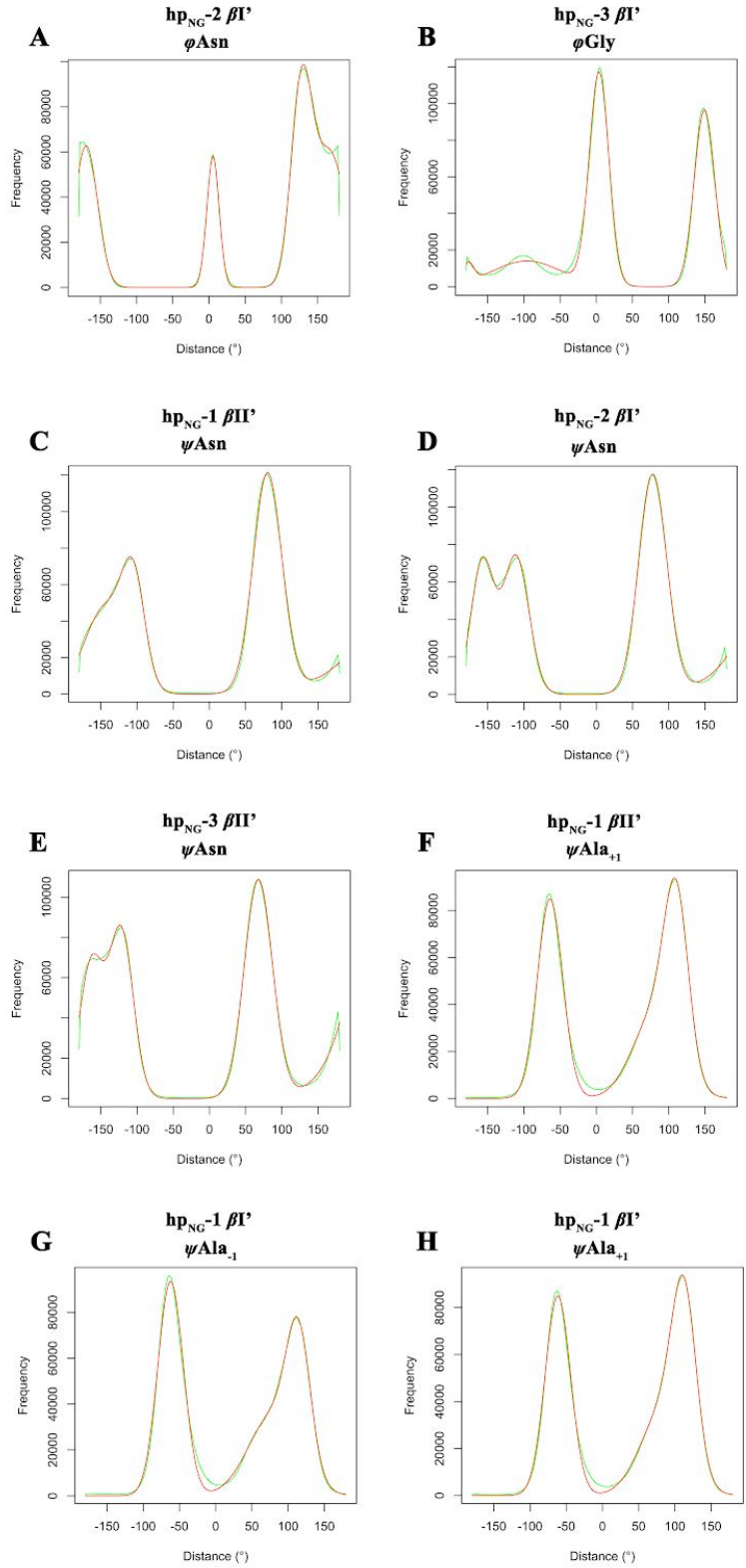

**Figure 4.** Histograms showing the distribution of distances (in °) between the *ab initio* φ,ψ values and the φ,ψ values obtained from the MD simulations for selected residues. While there are torsion angles of specific residues that fluctuate around the *ab initio* values ($d=0°$) during the simulation (A-B), some torsions present a significant divergence from the proposed values (C-H).

**Figure 5**

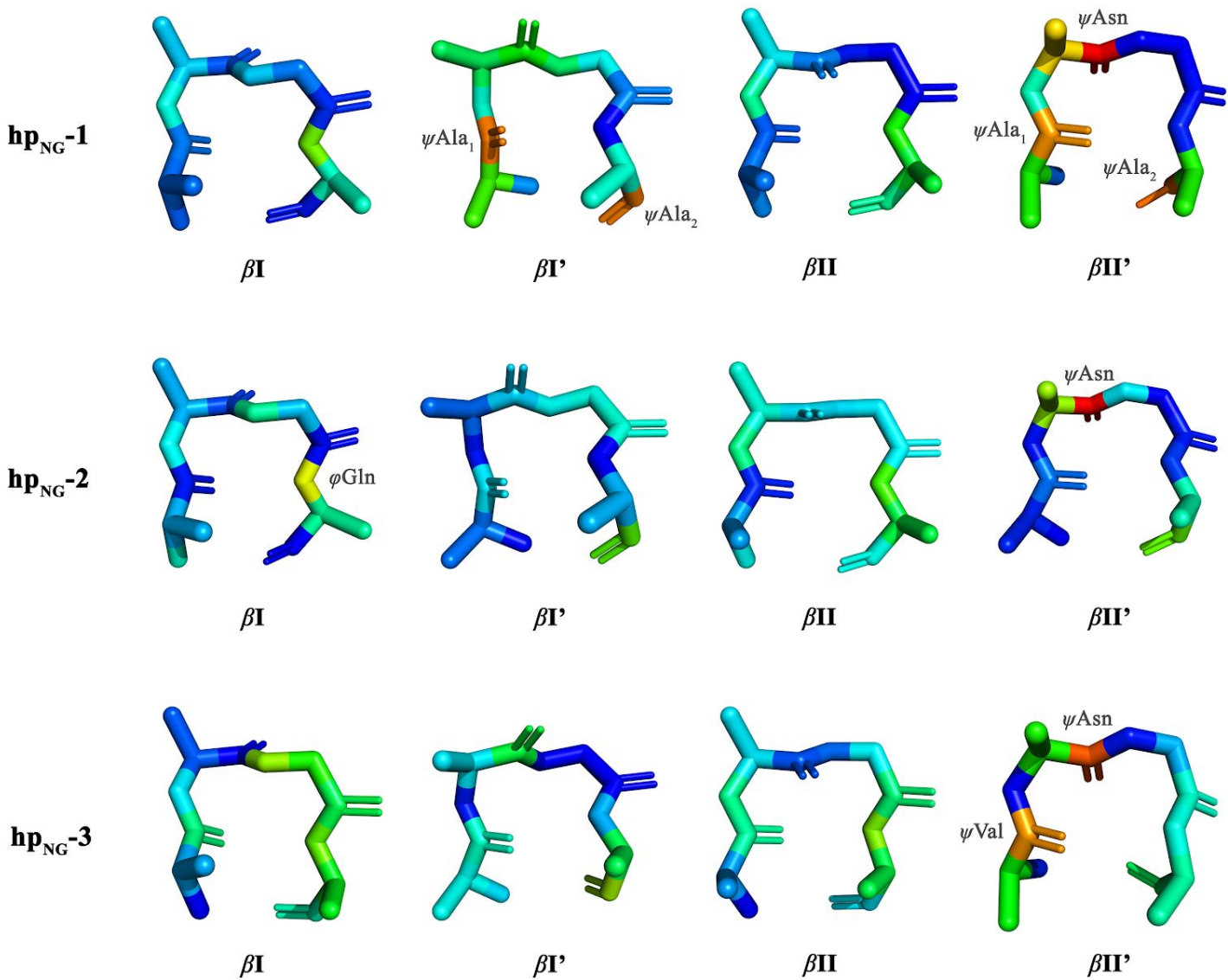

**Figure 5.** Shown are the *ab initio* structures illustrating the torsion angle discrepancies in relation with the MD derived angles. The colour coding is indicative of the mean distance of the MD $\varphi,\psi$ values from the corresponding *ab initio* ones, varying from blue (small deviations) to green (moderate deviations) and red (large deviations). Shown with a label, are the torsion angles presenting the greatest deviation through the simulations.

# On the propensity of Asn-Gly-containing heptapeptides to form *β*-turn structures : comparison between ab initio quantum mechanical calculations and Molecular Dynamics simulations.

Dimitrios A. Mitsikas & Nicholas M. Glykos*

*Department of Molecular Biology and Genetics, Democritus University of Thrace, University campus, 68100 Alexandroupolis, Greece, Tel +30-25510-30620, Fax +30-25510-30620, https://utopia.duth.gr/glykos/ , glykos@mbg.duth.gr*

**Table SI**

**Table SI.** Structural analysis and *β*-turns occupancy in the dPCA derived clusters[a]

| | No. of Cluster | $P_i$ (%) | X$_{-1}$-Asn | Asn-Gly | Gly-X$_{+1}$ |
|---|---|---|---|---|---|
| | 1 | 11.88 | IV (0.1%) | - | **II' (4.0%)** / IV (1.4%) |
| | 2 | 4.26 | - | **I (2.7%)** / IV (0.1%) | I (1.1%) / IV (1.1%) |
| | 3 | 8.65 | I (2.4%) / IV (1.5%) | **I (3.1%)** / IV (1.1%) | I (1.9%) / VIII (0.1%) / IV (1.2%) |
| | 4 | 7.23 | - | - | II (0.2%) / **IV (0.3%)** |
| | 5 | 7.67 | **I (5.0%)** / IV (2.0%) | IV (0.2%) | I' (0.1%) / IV (0.6%) |
| | 6 | 6.32 | VIII (0.4%) / IV (0.3%) | - | **II' (1.5%)** / IV (0.8%) |
| | 7 | 5.16 | **VIII (0.4%)** / IV (0.3%) | - | II (0.2%) / IV (0.3%) |
| | 8 | 10.74 | - | - | **I (4.6%)** / VIII (0.2%) / IV (2.3%) |
| hp$_{NG}$-1 | 9 | 9.40 | - | **II (6.6%)** / IV (1.5%) | I' (0.1%) / IV (1.7%) |
| | 10 | 5.10 | **I (2.6%)** / IV (1.9%) | IV (0.3%) | II (0.1%) / IV (0.2%) |
| | 11 | 6.64 | **I (3.6%)** / IV (2.1%) | - | II' (1.1%) / IV (0.6%) |
| | 12 | 5.97 | VIII (0.7%) / IV (0.3%) | - | **I (2.5%)** / VIII (0.2%) / IV (1.2%) |
| | 13 | 2.30 | - | **IV (0.2%)** | II (0.1%) / IV (0.1%) |
| | 14 | 2.75 | VIII (0.2%) / IV (0.1%) | **II (1.3%)** / IV (0.6%) | IV (0.5%) |
| | 15 | 3.01 | II (0.1%) / IV (1.0%) | **I' (2.0%)** | IV (0.2%) |
| | 16 | 2.06 | - | - | **II' (0.3%)** / IV (0.2%) |
| | 17 | 0.67 | - | - | **IV (0.1%)** |
| | 18 | 0.37 | **I' (0.1%) / IV (0.1%)** | I' (0.1%) | - |
| | 1 | 20.10 | II (0.7%) / IV (8.2%) | **I' (9.9%)** / IV (0.3%) | IV (6.8%) |
| | 2 | 14.09 | I (3.8%) / IV (2.6%) | **I (4.3%)** / IV (2.3%) | I (2.7%) / VIII (0.1%) / IV (2.4%) |
| | 3 | 4.55 | - | **I (2.4%)** / IV (0.1%) | I (1.0%) / IV (1.2%) |
| | 4 | 9.54 | **I (6.9%)** / IV (1.9%) | IV (0.4%) | I' (0.1%) / IV (0.7%) |
| | 5 | 10.02 | - | - | **I (5.1%)** / VIII (0.2%) / IV (2.6%) |
| | 6 | 10.13 | VIII (0.2%) / IV (0.2%) | - | **II' (5.8%)** / IV (0.9%) |
| | 7 | 3.24 | **I (1.7%)** / IV (1.3%) | IV (0.3%) | IV (0.2%) |
| hp$_{NG}$-2 | 8 | 5.98 | VIII (0.4%) / IV (0.3%) | - | **I (2.6%)** / VIII (0.2%) / IV (1.4%) |
| | 9 | 7.55 | - | **II (4.6%)** / IV (1.2%) | IV (2.4%) |
| | 10 | 4.39 | VIII (0.1%) / IV (0.2%) | **II (1.8%)** / IV (1.1%) | IV (0.6%) |
| | 11 | 2.42 | **VIII (0.2%) / IV (0.2%)** | - | IV (0.1%) |
| | 12 | 2.41 | - | - | **II (0.1%) / IV (0.1%)** |
| | 13 | 2.85 | **I (1.9%)** / IV (0.7%) | - | II' (0.3%) / IV (0.2%) |
| | 14 | 2.44 | - | - | **II' (0.3%)** / IV (0.2%) |
| | 15 | 0.28 | - | - | - |
| | 1 | 20.66 | I (3.8%) / IV (2.4%) | **I (6.2%)** / IV (1.6%) | I (2.9%) / VIII (0.1%) / IV (3.2%) |
| | 2 | 17.38 | **I (14.0%)** / IV (2.7%) | IV (0.5%) | I' (0.1%) / IV (0.8%) |
| | 3 | 16.50 | II (0.9%) / IV (7.7%) | **I' (9.9%)** / IV (0.1%) | IV (5.2%) |
| | 4 | 4.22 | - | **I (2.3%)** | I (0.9%) / IV (1.1%) |
| | 5 | 11.33 | - | - | **I (6.0%)** / VIII (0.2%) / IV (3.1%) |
| | 6 | 6.92 | - | **II (3.4%)** / IV (1.3%) | IV (1.9%) |
| | 7 | 7.48 | VIII (0.1%) / IV (0.1%) | **II (4.8%)** / IV (1.2%) | IV (2.4%) |
| hp$_{NG}$-3 | 8 | 3.87 | VIII (0.4%) / IV (0.2%) | - | **I (1.7%)** / VIII (0.1%) / IV (1.0%) |
| | 9 | 2.65 | - | - | **II' (0.3%)** / IV (0.3%) |
| | 10 | 2.71 | - | - | **IV (0.1%)** |
| | 11 | 1.40 | I (1.0%) / **IV (0.3%)** | IV (0.1%) | - |
| | 12 | 1.74 | **I (1.2%)** / IV (0.4%) | - | II' (0.1%) / IV (0.1%) |
| | 13 | 1.64 | **VIII (0.1%)** / IV (0.1%) | - | **IV (0.1%)** |
| | 14 | 1.08 | **VIII (0.1%)** / IV (0.1%) | - | **II' (0.1%)** / IV (0.1%) |
| | 15 | 0.41 | - | - | - |

[a]From left to right are listed the number of cluster, its population probability in (%), and the occupancy of different turn types in each cluster regarding different combinations of $i+1$ and $i+2$ turn residues from the peptides' central four-residue part. The assignment of $β$-turns was performed using the PROMOTIF program. Types VIa1, VIa2 and VIb are excluded from our analysis as they require a Pro residue in position $i+2$. The most populated turn type in each cluster is highlighted with bold.

**Table SII**

**Table SII.** Z-scores and corresponding probabilities for the observed deviations between the *ab initio* and the MD-derived torsion angles

|  | hp_NG-1 | | | hp_NG-2 | | | hp_NG-3 | | |
|---|---|---|---|---|---|---|---|---|---|
|  | Torsion angle | Z-score | Probability | Torsion angle | Z-score | Probability | Torsion angle | Z-score | Probability |
| βI | φAla | -0.63 | 0.53 | φVal | -1.33 | 0.18 | φVal | 0.06 | 0.95 |
|  | ψAla | -0.62 | 0.54 | ψVal | 0.49 | 0.62 | ψVal | -1.36 | 0.17 |
|  | φAsn | -1.25 | 0.21 | φAsn | -1.12 | 0.26 | φAsn | -1.11 | 0.27 |
|  | ψAsn | 0.60 | 0.55 | ψAsn | 0.52 | 0.60 | ψAsn | -0.10 | 0.92 |
|  | φGly | 1.04 | 0.30 | φGly | 0.31 | 0.76 | φGly | 2.98 | 0.00 |
|  | ψGly | -0.25 | 0.80 | ψGly | 1.02 | 0.31 | ψGly | -1.18 | 0.24 |
|  | φAla | 1.81 | 0.70 | φGln | 2.53 | 0.01 | φLeu | 1.89 | 0.06 |
|  | ψAla | 0.30 | 0.76 | ψGln | -0.14 | 0.89 | ψLeu | 0.12 | 0.90 |
| βI' | φAla | 0.78 | 0.44 | φVal | 0.29 | 0.77 | φVal | 0.99 | 0.32 |
|  | ψAla | 3.52 | 0.00 | ψVal | 1.14 | 0.25 | ψVal | 1.20 | 0.23 |
|  | φAsn | -1.32 | 0.19 | φAsn | -0.66 | 0.51 | φAsn | -0.64 | 0.52 |
|  | ψAsn | -0.99 | 0.32 | ψAsn | -0.77 | 0.44 | ψAsn | -1.07 | 0.28 |
|  | φGly | -1.84 | 0.06 | φGly | -1.42 | 0.16 | φGly | -0.29 | 0.77 |
|  | ψGly | 0.41 | 0.68 | ψGly | 0.90 | 0.37 | ψGly | 0.08 | 0.94 |
|  | φAla | 0.21 | 0.83 | φGln | 0.22 | 0.82 | φLeu | -0.66 | 0.51 |
|  | ψAla | 3.51 | 0.00 | ψGln | 2.12 | 0.03 | ψLeu | -2.26 | 0.02 |
| βII | φAla | -0.76 | 0.45 | φVal | -1.17 | 0.24 | φVal | 0.04 | 0.97 |
|  | ψAla | -0.53 | 0.60 | ψVal | -0.38 | 0.70 | ψVal | -1.29 | 0.20 |
|  | φAsn | -1.50 | 0.13 | φAsn | -1.53 | 0.13 | φAsn | -1.62 | 0.10 |
|  | ψAsn | 0.55 | 0.58 | ψAsn | 0.83 | 0.41 | ψAsn | 0.45 | 0.65 |
|  | φGly | -0.39 | 0.70 | φGly | -1.33 | 0.18 | φGly | 0.76 | 0.45 |
|  | ψGly | 0.06 | 0.95 | ψGly | 0.68 | 0.50 | ψGly | -1.07 | 0.28 |
|  | φAla | 1.57 | 0.12 | φGln | 2.03 | 0.04 | φLeu | 1.97 | 0.05 |
|  | ψAla | 1.40 | 0.16 | ψGln | 1.09 | 0.28 | ψLeu | 0.86 | 0.39 |
| βII' | φAla | 0.67 | 0.50 | φVal | 0.14 | 0.89 | φVal | 0.40 | 0.69 |
|  | ψAla | 3.31 | 0.00 | ψVal | 1.18 | 0.24 | ψVal | 3.13 | 0.00 |
|  | φAsn | -1.41 | 0.16 | φAsn | -0.58 | 0.56 | φAsn | -0.16 | 0.87 |
|  | ψAsn | -4.06 | 0.00 | ψAsn | -4.02 | 0.00 | ψAsn | -3.46 | 0.00 |
|  | φGly | -0.04 | 0.97 | φGly | 1.24 | 0.21 | φGly | -0.37 | 0.71 |
|  | ψGly | 0.16 | 0.87 | ψGly | 0.23 | 0.82 | ψGly | 0.90 | 0.37 |
|  | φAla | 0.39 | 0.70 | φGln | -0.46 | 0.64 | φLeu | -1.31 | 0.19 |
|  | ψAla | 3.64 | 0.00 | ψGln | 2.14 | 0.03 | ψLeu | -0.14 | 0.89 |

**Figure S1**

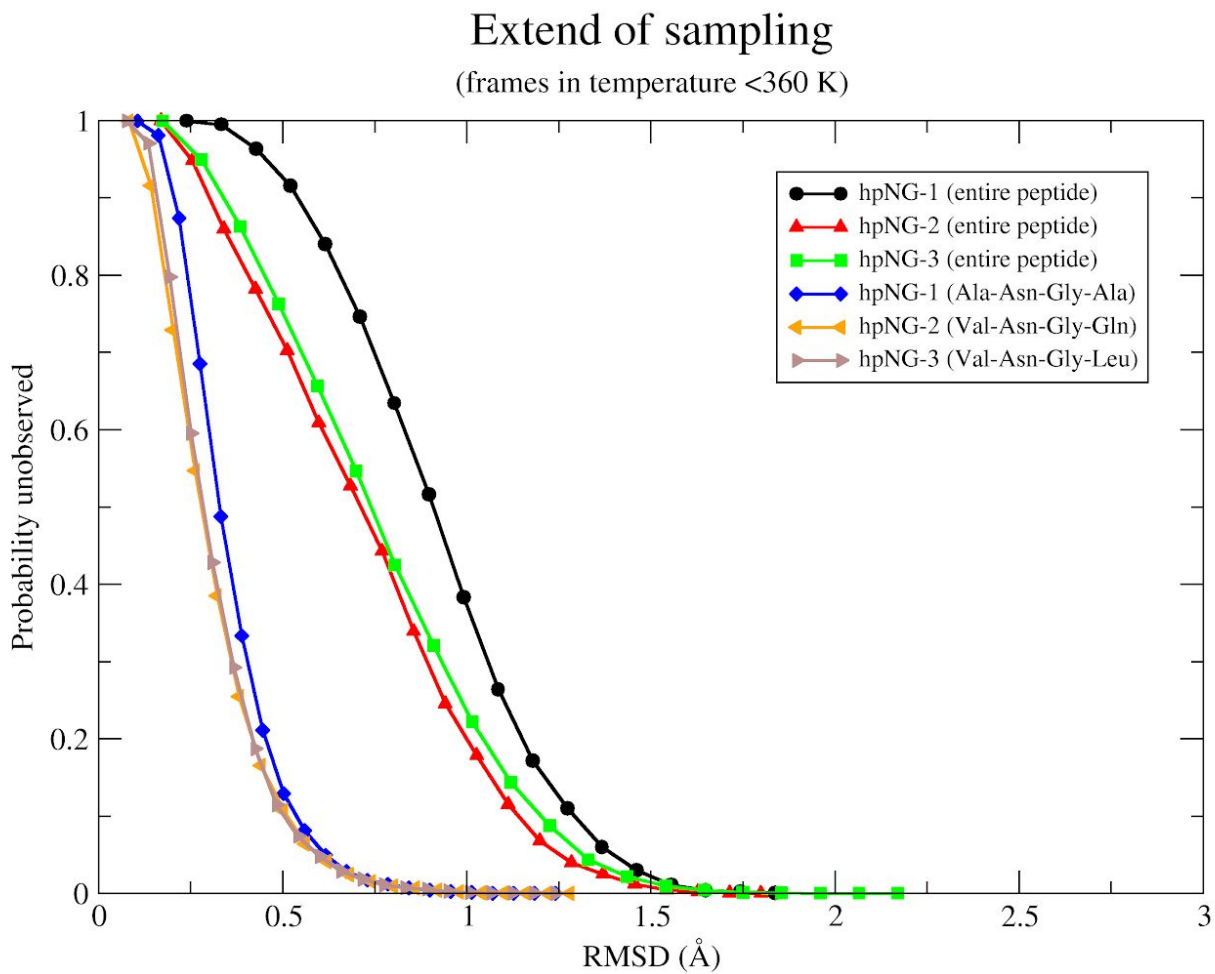

**Figure S1.** Extent of sampling and statistical significance. Results from the application of Good-Turing statistics to the three trajectories, for both the full-length peptides and their four-residue central part, obtained using only structures associated with temperatures ≤ 360 K.

**Figure S2**

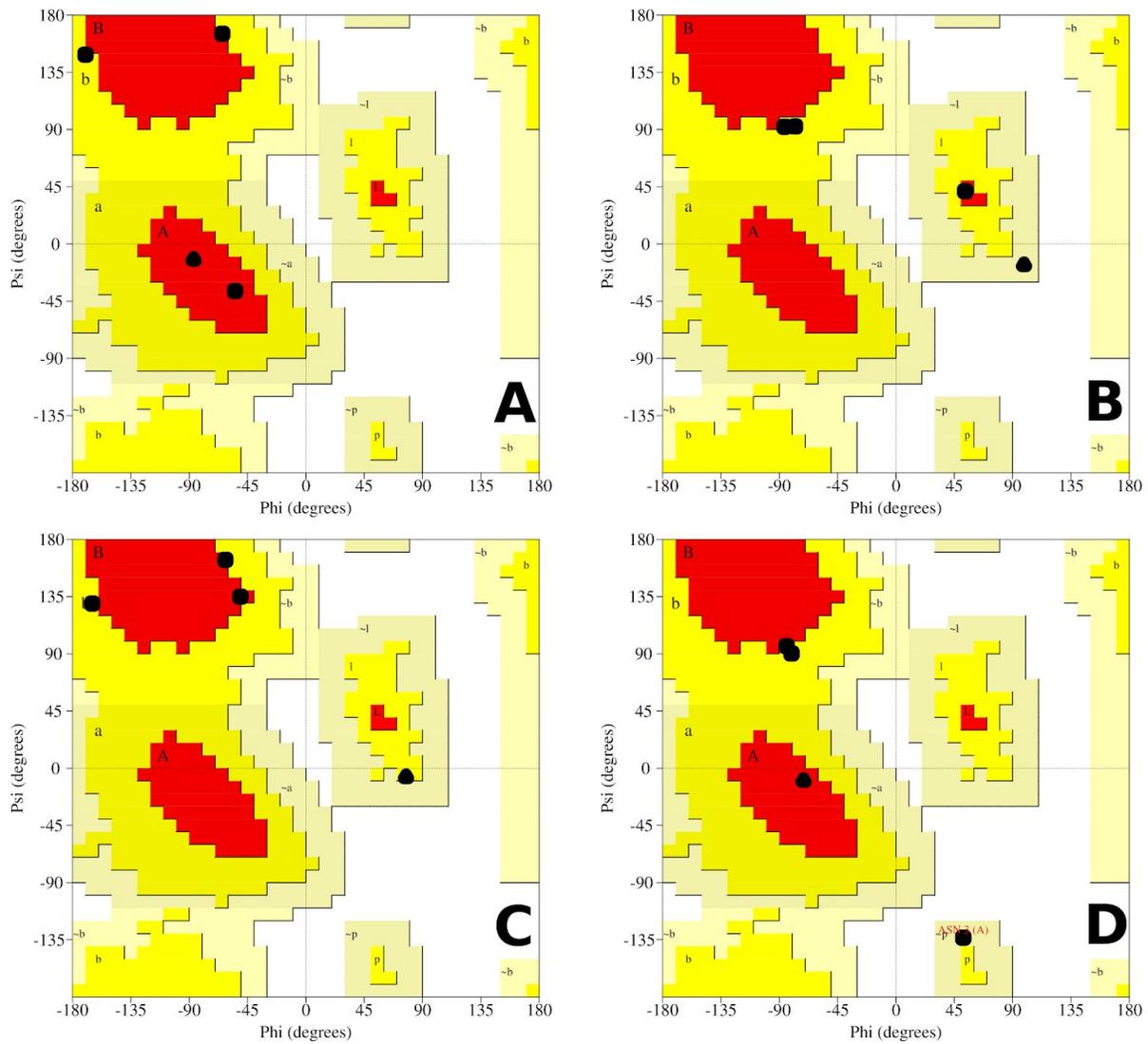

**Figure S2.** Ramachandran plots showing the *φ,ψ* torsion angle values of the four central residues of hp$_{NG}$-1 for (A) *β*I, (B) *β*I', (C) *β*II and (D) *β*II' *ab initio* turn types. Areas in red, yellow, beige and white represent the core, the allowed, the generous and the disallowed regions respectively. Non-glycine residues are presented here with black square signs and glycine residues with black triangles. Figures were generated using PROCHECK.

**Figure S3**

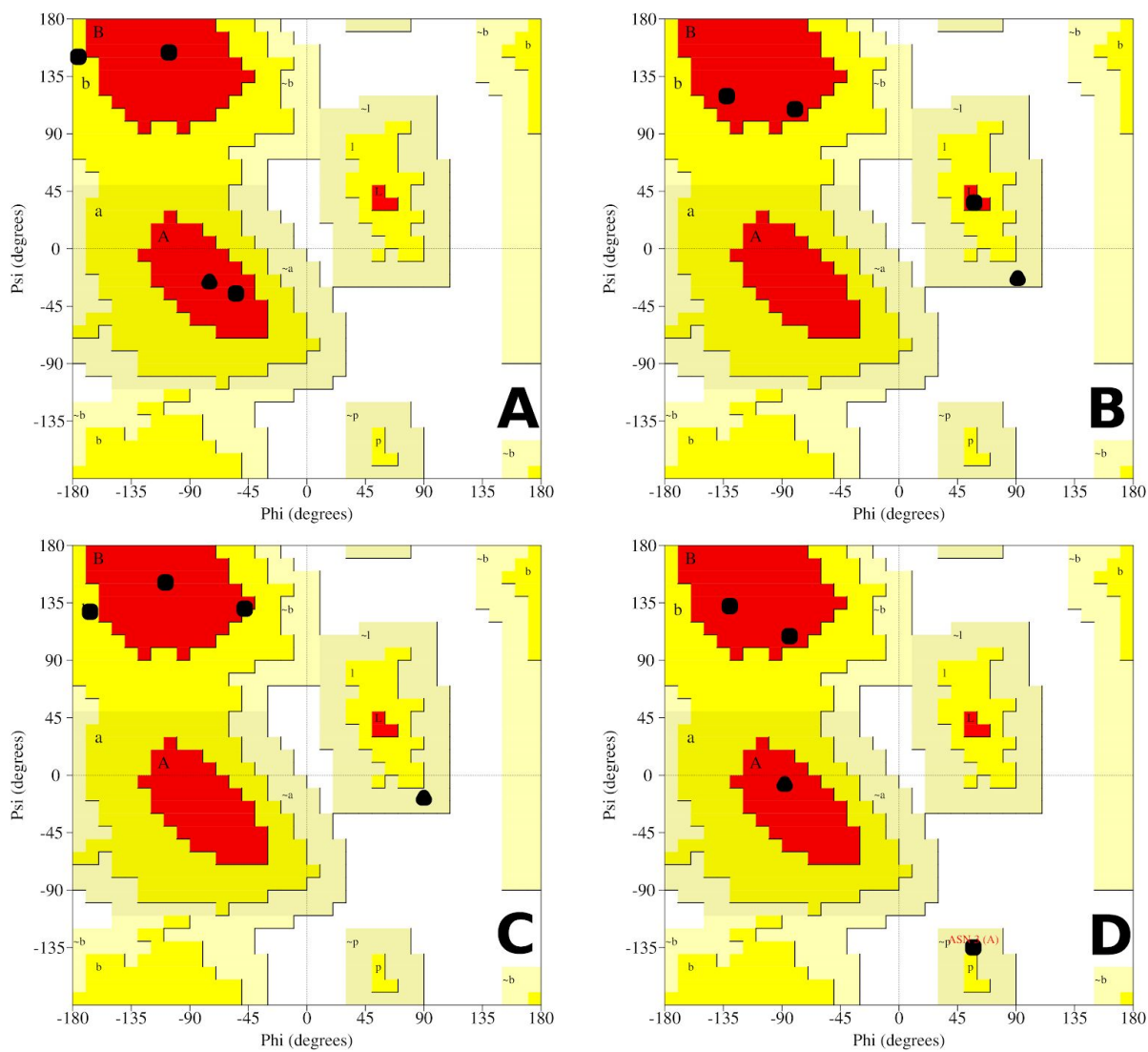

**Figure S3.** Ramachandran plots showing the *φ,ψ* torsion angle values of the four central residues of hp$_{NG}$-2, for (A) *β*I, (B) *β*I', (C) *β*II and (D) *β*II' *ab initio* turn types. Areas in red, yellow, beige and white represent the core, the allowed, the generous and the disallowed regions respectively. Non-glycine residues are presented here with black square signs and glycine residues with black triangles. Figures were generated using PROCHECK.

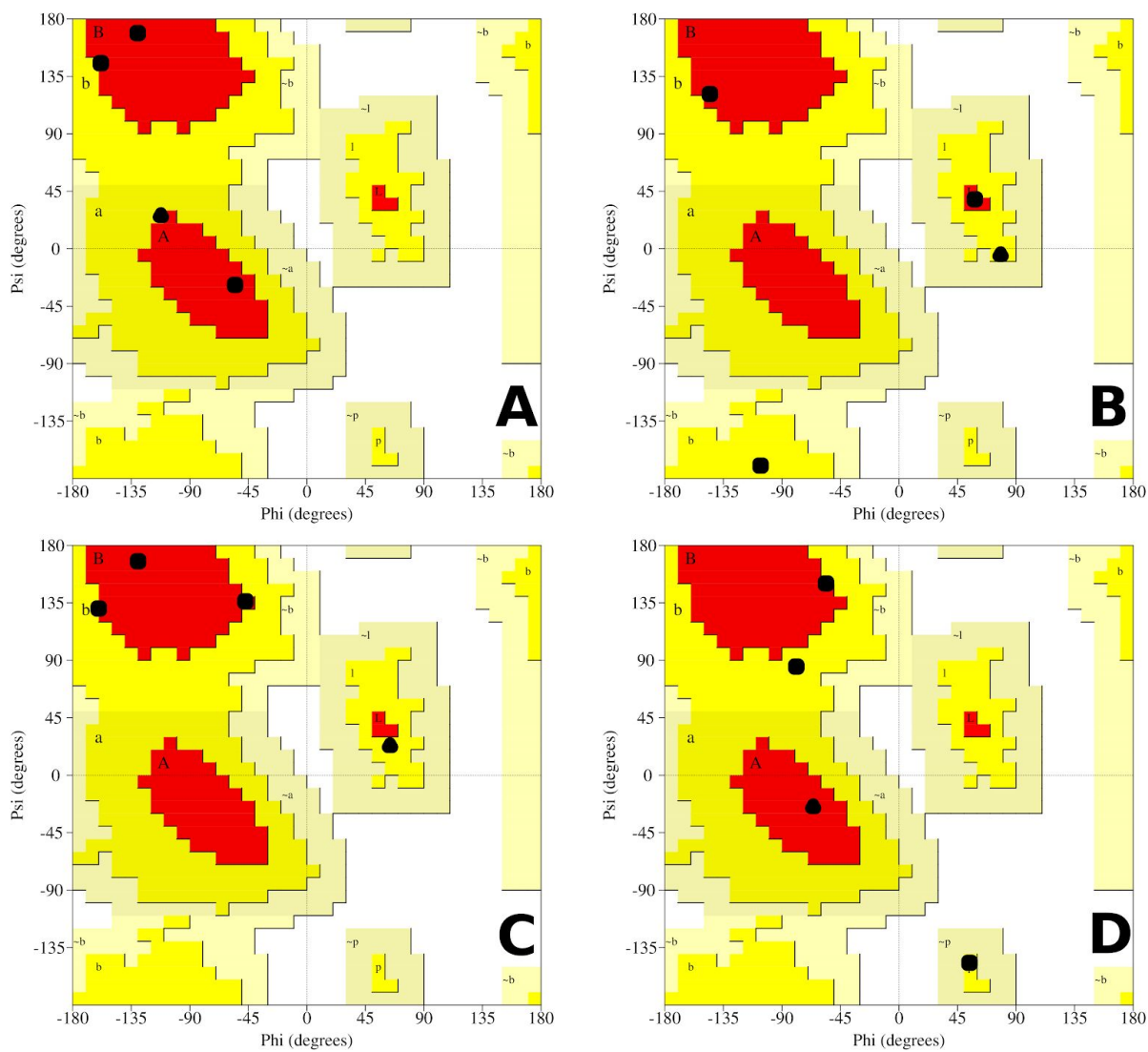

**Figure S4.** Ramachandran plots showing the $\varphi,\psi$ torsion angle values of the four central residues of hp$_{NG}$-3, for (A) $\beta$I, (B) $\beta$I', (C) $\beta$II and (D) $\beta$II' *ab initio* turn types. Areas in red, yellow, beige and white represent the core, the allowed, the generous and the disallowed regions respectively. Non-glycine residues are presented here with black square signs and glycine residues with black triangles. Figures were generated using PROCHECK.

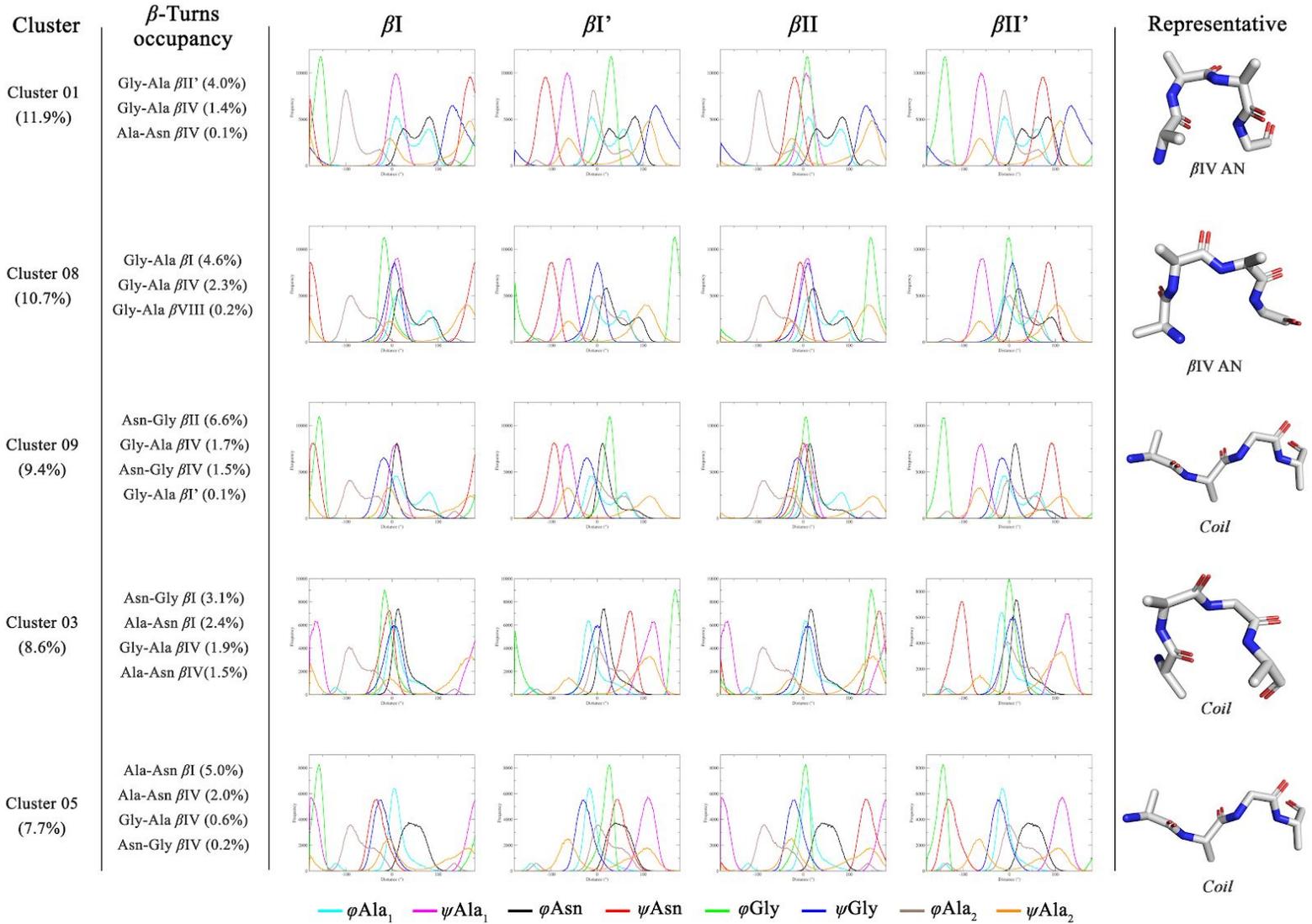

**Figure S5.** Results from the torsion angle analysis of the five most prominent dPCA clusters of hp$_{NG}$-1 for every turn type. Shown from left to right is the number of the cluster along with its population, the occupancy of *β*-turn motifs in each cluster, the histograms showing the distribution of distances (in degrees) between the *ab initio* *φ,ψ* values of the 4-residue central part and the respective *φ,ψ* values obtained from the simulation, and the corresponding backbone representative structures of each cluster. Representatives with random coil conformation are depicted here only with their Ala-Asn-Gly-Ala part.

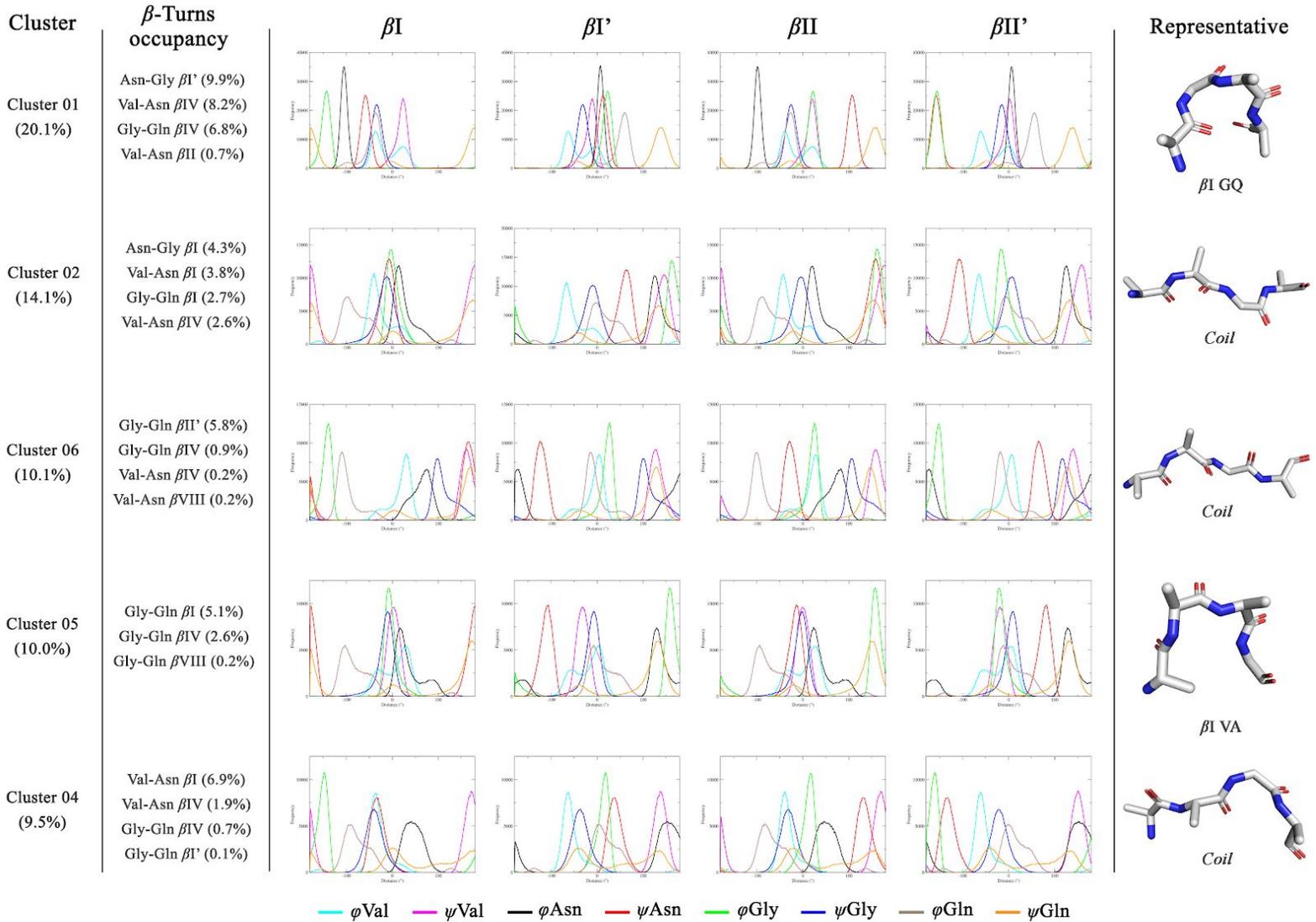

**Figure S6.** Results from the torsion angle analysis of the five most prominent dPCA clusters of hp$_{NG}$-2 for every turn type. Shown from left to right is the number of the cluster along with its population, the occupancy of β-turn motifs in each cluster, the histograms showing the distribution of distances (in degrees) between the *ab initio* φ,ψ values of the 4-residue central part and the respective φ,ψ values obtained from the simulation, and the corresponding backbone representative structures of each cluster. Representatives with random coil conformation are depicted here only with their Val-Asn-Gly-Gln part.

**Figure S7**

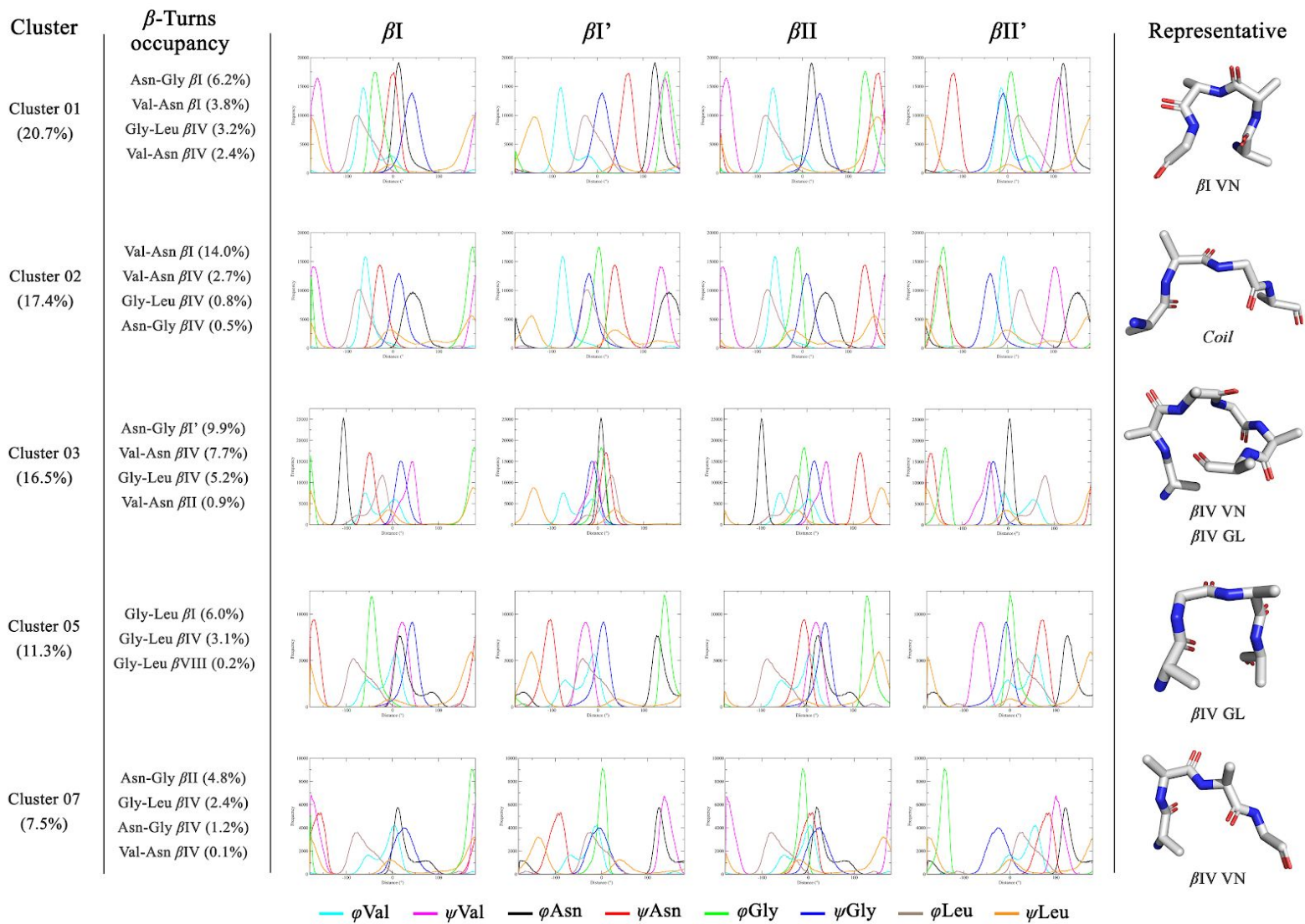

**Figure S7.** Results from the torsion angle analysis of the five most prominent dPCA clusters of hp$_{NG}$-3 for every turn type. Shown from left to right is the number of the cluster along with its population, the occupancy of β-turn motifs in each cluster, the histograms showing the distribution of distances (in degrees) between the *ab initio* φ,ψ values of the 4-residue central part and the respective φ,ψ values obtained from the simulation, and the corresponding backbone representative structures of each cluster. Representatives with random coil conformation are depicted here only with their Val-Asn-Gly-Leu part.

**Figure S8**

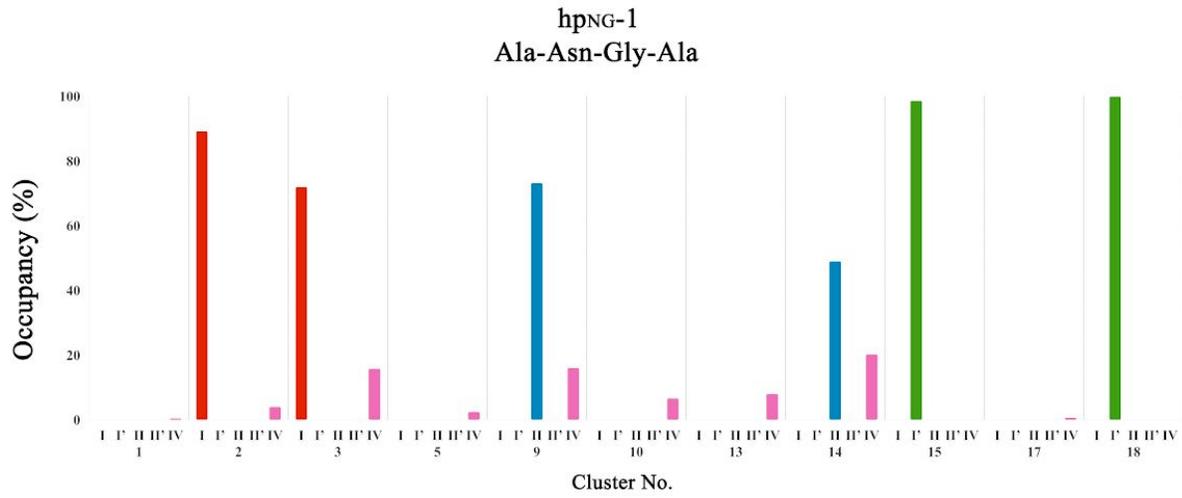

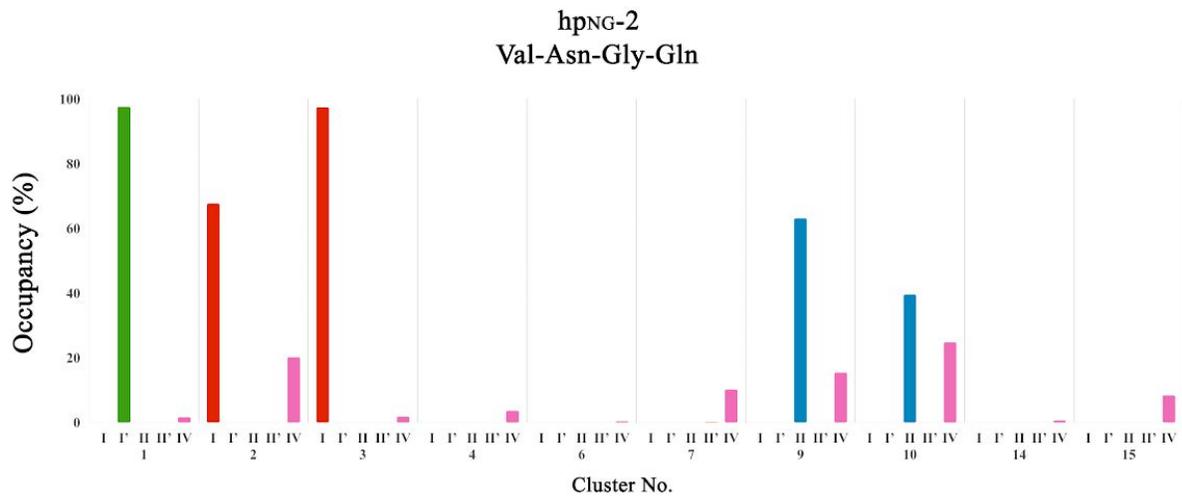

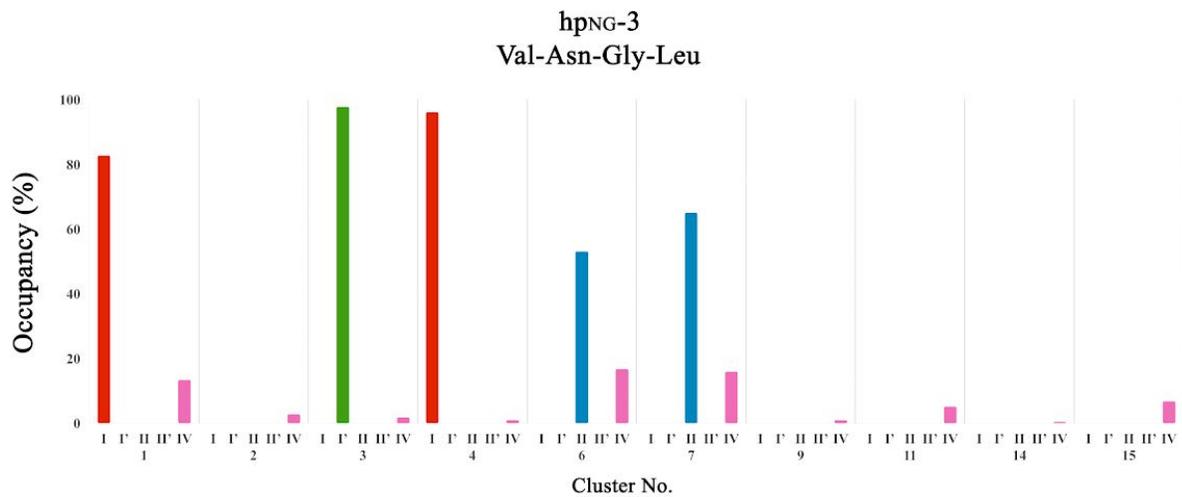

**Figure S8.** The above charts show the occupancy of Asn-Gly *β*-turns among a set of 500 equally spaced structures from every cluster of the three heptapeptides, as obtained by the dihedral angle Principal Component Analysis. Clusters in which *β*-turns were not identified among the set of structures are omitted. Shown at the horizontal axis are clusters' numbers and the different turn types. The vertical axis shows the % occupancy of the different turn types in each cluster. For the analysis, only the four central residues of each structure were taken into account.